\documentclass[11pt]{article}
\pdfoutput=1
\usepackage[dvipsnames]{xcolor}
\usepackage[all]{xy}
\usepackage{amsmath}
\usepackage{amssymb}
\usepackage{braket}
\usepackage{caption}
\usepackage{float}
\restylefloat{table}
\usepackage{fancyhdr}
\newcommand*\rfrac[2]{{}^{#1}\!/_{#2}}
\usepackage{hyperref}
\hypersetup{
    unicode=false,          
    pdftoolbar=true,        
    pdfmenubar=true,        
    pdffitwindow=false,     
    pdfstartview={FitH},    
    pdftitle={My title},    
    pdfauthor={Author},     
    pdfsubject={Subject},   
    pdfcreator={Creator},   
    pdfproducer={Producer}, 
    pdfkeywords={keywords}, 
    pdfnewwindow=true,      
    colorlinks=false,       
    linkcolor=red,          
    citecolor=cayn,        
    filecolor=magenta,      
    urlcolor=cyan           
}
 
\newcommand{\xA}{{\mbox{\footnotesize(1)}}}
\newcommand{\xB}{{\mbox{\footnotesize(2)}}}
\newcommand{\delt}{\delta\,(\,{\mbox{\footnotesize{2}}}\,-\,{\mbox{\footnotesize 1}}\,)}
\newcommand{\deltP}{\delta^{\prime}\,(\,{\mbox{\footnotesize{2}}}\,-\,{\mbox{\footnotesize 1}}\,)}
\newcommand{\defeq}{\mathrel{\mathop:}=}
\newcommand{\eqdef}{\mathrel{\mathop=}:}
\newcommand{\sN}{\mathcal{N}}
\newcommand{\As}{{\scriptsize\mathcal{A}}}
\newcommand{\Bs}{{\scriptsize\mathcal{B}}}
\newcommand{\Cs}{{\scriptsize\mathcal{C}}}
\newcommand{\Ds}{{\scriptsize\mathcal{D}}}
\newcommand{\Es}{{\scriptsize\mathcal{E}}}

\newcommand{\Ms}{{\scriptsize\mathcal{M}}}
\newcommand{\Ns}{{\scriptsize\mathcal{N}}}
\newcommand{\Os}{{\scriptsize\mathcal{O}}}

\newcommand{\Oo}{{\scriptsize\mathcal O}}
\newcommand{\Eone}{E^{\scriptstyle (1)}}

\def\be{\begin{equation}}
\def\ee{\end{equation}}

\def\ba{\begin{eqnarray}}
\def\ea{\end{eqnarray}}

\def\frac#1#2{{\textstyle{#1\over#2}}}
\def\Fontamici#1{{$\mathcal{#1}$}}
\def\go#1{{\mbox{{\scriptsize\Fontamici{#1}}}}}

\def\rp{{\color{red} . }}

\def\rgboo#1{\pdfliteral{#1 rg #1 RG}}
\def\pdfklink#1#2{%
	\noindent\pdfstartlink user
		{/Subtype /Link
		/Border [ 0 0 0 ]
		/A << /S /URI /URI (#2) >>}{\rgb{1 0 0}{#1}}%
	\pdfendlink}
\def\rgbo#1#2{\rgboo{#1}#2\rgboo{0 0 0}}
\def\rgb#1#2{\mark{#1}\rgbo{#1}{#2}\mark{0 0 0}}

\def\xxxlink#1{\pdfklink{[arXiv:#1]}{http://arXiv.org/abs/#1}}

\def\Gamma{\mathchar"0100}
\def\Delta{\mathchar"0101}
\def\Theta{\mathchar"0102}
\def\Lambda{\mathchar"0103}
\def\Xi{\mathchar"0104}
\def\Pi{\mathchar"0105}
\def\Sigma{\mathchar"0106}
\def\Upsilon{\mathchar"0107}
\def\Phi{\mathchar"0108}
\def\Psi{\mathchar"0109}
\def\Omega{\mathchar"010A}

\newenvironment{rcases}
  {\left.\begin{aligned}}
  {\end{aligned}\right\rbrace}

\begin{document}

\title{{\bf\color{blue} T-duality off shell in 3D Type II superspace}\\[-2in]
{\normalsize March 26, 2014\hfill YITP-SB-14-9}\\[1.8in]}
\date{}
\author{Martin Pol\'a\v cek\footnote{\pdfklink{martin.polacek@stonybrook.edu}{mailto:martin.polacek@stonybrook.edu}},\ \ 
Warren Siegel\footnote{\pdfklink{siegel@insti.physics.sunysb.edu}{mailto:siegel@insti.physics.sunysb.edu},
	\pdfklink{http://insti.physics.sunysb.edu/\~{}siegel/plan.html}{http://insti.physics.sunysb.edu/\%7Esiegel/plan.html}}\\
{\textit{C. N. Yang Institute for Theoretical Physics}}\\ {\textit{State University of New York, Stony Brook, NY 11794-3840}}}

\maketitle

\begin{abstract}

\normalsize
We give the manifestly T-dual formulation of the massless sector of the classical 3D Type II superstring in off-shell 3D $\sN\,=\,2$ superspace, including the action.  It has a simple relation to the known superspace of 4D $\sN\,=\,1$ supergravity in 4D M-theory via 5D F-theory.  The pre potential appears as part of the vielbein, without derivatives.

\end{abstract}

\newpage

\section{Introduction}
In the paper \cite{natural} we have discovered that the curvature tensor previously discovered in \cite{warren} could be obtained in a manifestly T-dual way. We have seen that using the techniques of the T-dually extended space-time and the coset construction we could naturally find geometrical objects (like the Riemann curvature as a part of the torsion) i.e. the structure of the physical theory. We would like to extend those techniques (coset construction, orthogonality constraints, etc.) to the supersymmetric case, i.e. to work directly with T-dually extended superspace. In this article (as a starting point for a bigger program on T-dually extended superspaces)  we consider the $3$ dimensional T-dually extended superspace. The higher dimensional case is discussed in \cite{warren1}. We would also see that this (toy) model of $3$ dimensional T-dually extended space goes with the idea of lower dimensional F-theory (i.e. lower dimensional analogue of the $12$ dimensional F-theory, see \cite{Vafa}). For simplicity we will work in the linearised regime. At the end we will show that the physical spectrum (and the structure) of the theory coincides with the $\sN=2$ supergravity in $3$ dimensions (after the compactification). That should be expected since as we will show the classical $\sN=1$ supergravity in $4$ dimensions could be interpreted as to have the same F-theory origin as the T-dual $3$D supergravity. So does the $3$D $\sN=2$ supergravity (after the compactification of $4$D $\sN=1$ supergravity to $3$D).

We are following the procedure described in the articles \cite{natural}, \cite{warren} and \cite{warren1}. The differences are that we are working just to linear order in fields and in the $3$D T-dual superspace. On top of that will also find the relation of the T-dually extended theory to the (lower dimensional analog of) F-theory.        
\section{F-theory (membrane vs. strings)}
\subsection{F-theory and its compactification}
The F-theory has first been proposed by Cumrun Vafa as $12$ dimensional theory, see \cite{Vafa}. The theory is further compactified on the two-torus or more generally on the elliptically fibered Calabi-Yau manifolds. We discuss the $5$ dimensional analogue of this theory. We want to motivate the natural identification between the  $4$D $\sN=1$ supergravity, further compactified to a $3$D $\sN=2$ (the $3$D $\sN=2$ supergravity is recently discussed in \cite{rozlucka}),  and the T-dual $3$ D $\sN=2$ string theory. Both can be thought to have an origin in higher dimensional F-theory. This theory will be further compactified in two ways. One compactification produces the $4$ dimensional M-theory that will effectively become the $\sN=1$ supergravity with the specific chiral compensator that contains a $3$-form. This is expected since this $\sN=1$ supergravity is an effective theory of 2-branes (discussion of the lower dimensional supersymmetric membrane theory could be found in \cite{martinec}, (super) membrane theory discussed in \cite{smutnySom2}, \cite{smutnySom3}, \cite{smutnySom4}, \cite{smutnySom5}). The other compactification gives the $3$ dimensional T-dual $\sN=2$ string theory so effectively the T-dual $\sN=2$ supergravity. 
\subsection{\texorpdfstring{$5$D vs. $4$D vs. $3$D - compactifications}{5D vs. 4D vs. 3D - compactifications} }    
 
The $5$ dimensional F-theory is the (supersymmetric) $2$-brane theory in the space with the signature $(\,+,\,+,\,+,\,-,\,-\,)$ . The Lorentz group is $SO\,(\,3,\,2\,)$. We can pick the time direction and compactify the F-theory along one time direction, so we will get the Lorentz group breaking $SO\,(\,3,\,2\,)\,\rightarrow\,SO(\,3,\,1\,)$. The $4$ dimensional $\sN=1$ $SO(\,3,\,1\,)$ theory is just the $4$ dimensional M-theory, which is effectively the $4$ dimensional $\sN=1$ supergravity. We can also pick the space direction and compactify the F-theory along this direction, so we will get: $SO\,(\,3,\,2\,)\,\rightarrow\,SO\,(\,2,\,2)\,\simeq\,SO\,(\,2,\,1\,)\,\otimes\,SO\,(\,2,\,1\,)$ what will become the T-dual $\sN=2$ string theory and effectively the T-dual $3$D $\sN=2$ supergravity.
If we further compactify the $4$ dimensional $\sN=1$ supergravity along the space direction we will get the $3$ dimensional $\sN=2$ supergravity coupled to a vector multiplet. On the other hand, if we take the T-dual $3$D $\sN=2$ theory and compactify half of the dimensions we would again get the $3$D $\sN=2$ supergravity coupled to a vector multiplet. We therefore have the natural identification of the objects from the $4$D $\sN=1$ supergravity (further compactified) and the T-dual $3$D $\sN=2$ supergravity. We can therefore use the techniques of T-dually extended superspace and derive the $3$D $\sN=2$ supergravity coupled to a vector multiplet. 
 
In the $4$D ($n=-\frac{1}{3}$ minimal and linearised) supergravity we have the pre potential $H_{\alpha\,\dot{\beta}}$ and the scalar pre potential $\mathcal{V}$. The scalar pre potential becomes a particular (chiral) compensator of the form $\phi\,=\,\bar{D}^2\,\mathcal{V}$. That contains a $3$-form, see section $4$.$4$.d in \cite{1001}, or more generally \cite{Gates}. This is expected since $4$D $\sN\,=\,1$ supergravity is the effective theory for $2$-branes. 

The $4$D $\sN=1$ gauge transformations are, see section $5$.$2$ in \cite{1001} or \cite{graph}:
\ba
\label{srncek}
\delta\,H_{\alpha\,\dot{\beta}}\,=\,D_{\alpha}\,\bar{L}_{\dot{\beta}}\,-\,\bar{D}_{\dot{\beta}}\,L_{\alpha}\,\,\,\,\text{and}\,\,\,\,\delta\,\mathcal{V}\,=\,D^{\alpha}\,L_{\alpha}\,+\,\bar{D}^{\dot{\alpha}}\,\bar{L}_{\dot{\alpha}}
\ea
where $D_{\alpha}$ and $\bar{D}_{\dot{\alpha}}$ are usual $4$D $\sN=1$ covariant derivatives. We can dimensionally reduce the theory to $3$D and obtain the $3$D $\sN=2$ theory. Using the dimensional reduction we get:
\ba
\label{srncek1}
D_{\alpha}\,=\,\frac{1}{\sqrt{2}}\,(\,D_{\alpha}\,+\,i\,D_{\alpha^{\prime}}\,)\,\,\,\,\text{and}\,\,\,\,\bar{D}_{\dot{\alpha}}\,=\,\frac{1}{\sqrt{2}}\,(\,D_{\alpha}\,-\,i\,D_{\alpha^{\prime}}\,)
\ea
where $D_{\alpha}$ and $D_{\alpha^{\prime}}$ are real 3D $\sN=2$ covariant derivatives. The gauge parameters can be written as: 
\ba
L_{\alpha}\,=\,\frac{1}{\sqrt{2}}\,(\,\Lambda_{\alpha}\,-\,i\,\Lambda_{\alpha^{\prime}}\,)\,\,\,\,\text{and}\,\,\,\,\bar{L}_{\dot{\alpha}}\,=\,\frac{1}{\sqrt{2}}\,(\,\Lambda_{\alpha}\,+\,i\,\Lambda_{\alpha^{\prime}}\,).
\ea
The $3$D $\sN=2$ gauge transformations thus are: 
\ba
\delta\,H_{(\alpha\,\dot{\beta})}&=&\delta\,H_{\alpha\,\beta^{\prime}}\,=\,i\,(\,D_{(\alpha^{\prime}}\,\Lambda_{\beta)}\,+\,D_{(\alpha}\,\Lambda_{\beta^{\prime})}\,)\\
\delta\,H_{[\alpha\,\dot{\beta}]}&=&\delta\,V\,=\,D^{\alpha}\,\Lambda_{\alpha}\,-\,D^{\alpha^{\prime}}\,\Lambda_{\alpha^{\prime}}\\
\delta\,\mathcal{V}&=&D^{\alpha}\,\Lambda_{\alpha}\,+\,D^{\alpha^{\prime}}\,\Lambda_{\alpha^{\prime}}
\ea 

The $4$D $\sN=1$ pre potential $H_{\alpha\,\dot{\beta}}\,\equiv\,(\,H_{(\alpha\,\dot{\beta})},\,H_{[\alpha\,\dot{\beta}]}\,)$ is a $4$D vector and becomes the $3$D vector $H_{(\alpha\,\beta^{\prime})}$ and a pre potential $V$ (for a vector multiplet). We also have the $4$D pre potential $\mathcal{V}$ (for the chiral compensator $\phi\,=\,\bar{D}^{2}\,\mathcal{V}$) that becomes the $3$D pre potential $\mathcal{V}$ . On the other hand the $3$D T-dual pre potential (symmetric part) $H_{(\alpha\,\beta^{\prime})}$ (after the dimensional reduction to  $3$D $\sN=2$) is again a vector (describes the conformal supergravity) but the $H_{[\alpha\,\beta^{\prime}]}$ becomes the pre potential $\mathcal{V}$, see the transformations (\ref{nunu}), and the pre potential $H_{\alpha\,{\beta}^{\prime}}$ is just part of vielbeins, see table (\ref{hrncek2}). Finally the $3$D T-dual $\sN=2$ pre potential $V$ becomes the pre potential for the vector multiplet in $3$D $\sN=2$ supergravity. 

Therefore we have an identifications between $3$D $\sN=2$ T-dual supergravity and $3$D $\sN=2$ supergravity coupled to a vector multiplet: $H_{(\alpha\,\beta^{\prime})}\,\rightarrow\,H_{(\alpha\,\beta^{\prime})},\,\,\,H_{[\alpha\,\beta^{\prime}]}\,\rightarrow\,\mathcal{V}\,\mbox{and}\,{V\,\rightarrow\,V}$.

We also have the identification between $4$D $\sN=1$ supergravity and $3$D $\sN=2$ supergravity coupled to a vector multiplet: $H_{(\alpha\,\dot{\beta})}\,\rightarrow\,H_{(\alpha\,\beta^{\prime})},\,\,\,H_{[\alpha\,\dot{\beta}]}\,\rightarrow\,V\,\mbox{and}\,{\mathcal{V}\,\rightarrow\,\mathcal{V}}$.

The situation could be summarised in the following diagram \ref{diagram}:
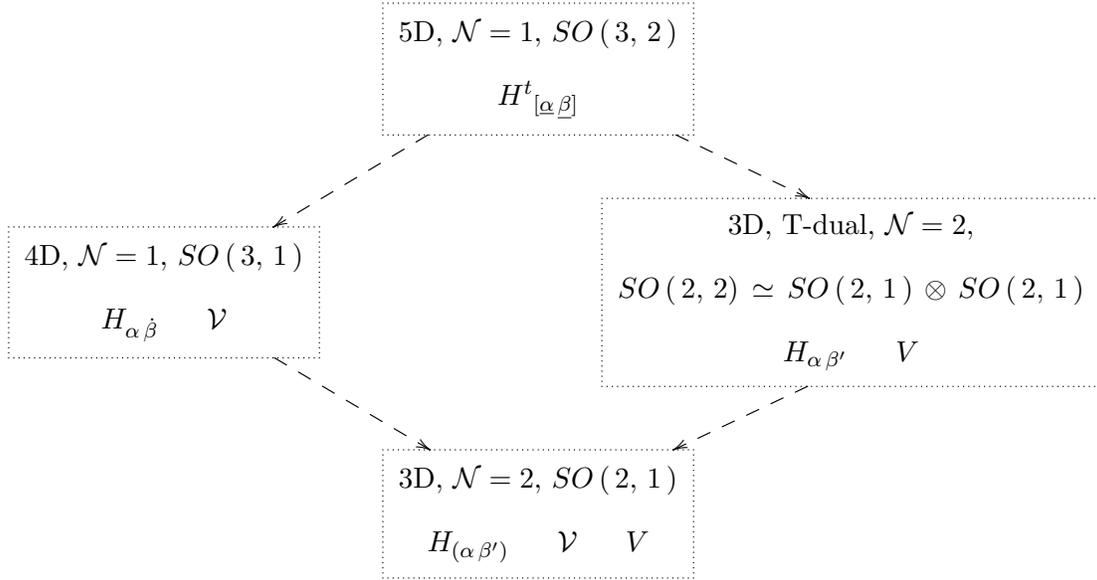
\begin{figure}[H]    
{\centering
\xymatrix{&&*++[F.]{\txt{$5$D, $\sN=1$,  $SO\,(\,3,\,2\,)$\\\\$H^{t}{}_{[\underline{\alpha}\,\underline{\beta}]}$}}\ar@{-->}[ld] \ar@{-->}[rd]&&\\
&*++[F.]{\txt{$4$D,  $\sN=1$,  $SO\,(\,3,\,1\,)$\\\\$H_{\alpha\,\dot{\beta}}\,\,\,\,\,\,\,\,\,\,{\mathcal{V}}$}}\ar@{-->}[rd]& & *++[l][F.]{\txt{$3$D,  T-dual,  $\sN=2$,\\\\$SO\,(\,2,\,2)\,\simeq\,SO\,(\,2,\,1\,)\,\otimes\,SO\,(\,2,\,1\,)$  \\\\$H_{\alpha\,\beta^{\prime}}\,\,\,\,\,\,\,\,\,\,V$}}\ar@{-->}[ld]&\\
&&*++[F.]{\txt{$3$D,  $\sN=2$,  $SO\,(\,2,\,1\,)$\\\\$H_{(\alpha\,\beta^{\prime})}\,\,\,\,\,\,\,\,\,\,{\mathcal{V}}\,\,\,\,\,\,\,\,\,\,V$}}&&
}
}
\caption{F-theory breaking \protect\label{diagram}}
\end{figure}
where $H^{t}{}_{[\underline{\alpha}\,\underline{\beta}]}$ is the $5$ dimensional pre potential ($\underline{\alpha}\,\in\,\{\,1,..4\,\}$,``$t$'' means that it is traceless, it has $5$ real components).

\section{Algebra}
We give very brief outline of the algebraic objects and steps that will lead to the formulation of the linearised T-dual $3$D supergravity. The interested reader may see the following references (where the subject is explained in great detail): \cite{natural}, \cite{warren}.
\subsection{Current algebra of \texorpdfstring{$Z_{\Ms}$}{Z_{\Ms}}}
As in the paper \cite{natural}, we consider the (super)string generalisation of the string oscillator algebra. Because of the T-duality and the (super)Bianchi identity the current algebra has a structure:
\be
\label{twotwo}
\mbox{[}\,Z_{\mbox{{\scriptsize\Fontamici{M}}}}\,\xA,\,Z_{\mbox{{\scriptsize\Fontamici{N}}}}\,\xB\,\mbox{]}\,=\,-i\,\eta_{\mbox{{\scriptsize\Fontamici{M}}}\,\mbox{{\scriptsize\Fontamici{N}}}}\,\deltP\,-\,i\,f_{\mbox{{\scriptsize\Fontamici{M}}}\,\mbox{{\scriptsize\Fontamici{N}}}}{}^{\mbox{{\scriptsize\Fontamici{P}}}}\,Z_{\,\mbox{{\scriptsize\Fontamici{P}}}}\,\delt
\ee
where $Z_{\Ms}\,\defeq\,(\,S_{MN},\,D_{\mu},\,P_{M},\,\Omega^{\mu},\,\Sigma^{MN}\,)$ is the generalisation of the (super)string oscillators and the metric $\eta_{\Ms\Ns}$ (given later). The $P_{M}$ generators are the $O\,(D,\,D)$ generalisation of string oscillators $P_{m}$. In the explicit $O(D,\,D)$ basis are the $P_{M}$ generators given as: $P_{M}\,\defeq\,(\,P_{m},\,X^{\prime\,m}\,)$. For the future purpose we want to use a different left/right basis. In left/right basis the $P_{M}\,\defeq\,(\,P_{\bf m},\,P_{\bf \tilde{m}}\,)\,=\,\frac{1}{\sqrt{2}}\,(\,P_{m}\,+\,X^{\prime}_{m},\,P_{m}\,-\,X^{\prime}_{m}\,)$. The Lorentz generators also have the left/right structure: $S_{MN}\,\defeq\,(\,S_{\bf mn},\,S_{\bf \tilde{m}\tilde{n}}),$ where $S_{\bf mn}$ are generators of left (or equivalently $S_{\bf \tilde{m}\tilde{n}}$ right) Lorentz transformations. The $D_{\mu}\,\defeq\,(\,D_{\bf \mu},\,D_{\bf \tilde{\mu}}\,)$ are the generators of left and right supersymmetry transformations. The generators $\Omega^{\mu}\,\defeq\,(\,\Omega^{\bf \mu},\,\Omega^{\bf \tilde{\mu}}\,)$ and $\Sigma^{MN}\,\defeq\,(\,\Sigma^{\bf mn},\,\Sigma^{\bf \tilde{m}\tilde{n}}\,)$ are the new generators, needed to satisfy the Bianchi identity. For further reference see \cite{natural}, \cite{warren}.
   
The full current algebra of $Z_{\Ms}$ oscillators (\ref{twotwo}) is the affine (super)Lie algebra (\ref{algebra}) and its explicit form is:
\begin{eqnarray}
\label{algebra}
{[}\,S_{\bf{m}\bf{n}}\,{\mbox{\footnotesize(1)}},\,S_{\bf{k}\bf{l}}\,{\mbox{\footnotesize(2)}}\,{]}&=
&-i\,\eta_{[\,\bf{m}\,[\,\bf{k}}\,S_{\bf{l}\,]\,\bf{n}\,]}\,\,\delta\,(\,{\mbox{\footnotesize{2}}}\,-\,{\mbox{\footnotesize 1}}\,)\\
{[}\,S_{\bf{m}\bf{n}}\,{\mbox{\footnotesize(1)}},\,D_{\bf{\rho}}\,\,{\mbox{\footnotesize(2)}}\,{]}&=&-\,i\,\frac{1}{2}\,(\,\gamma_{\bf{m}\bf{n}}\,)^{\bf{\sigma}}_{\bf{\rho}}\,D_{\bf{\sigma}}\,\delt\nonumber\\
{[}\,S_{\bf{m}\bf{n}}\,\xA,\,P_{\bf{k}}\,\xB\,{]}&=&i\,\eta_{\bf{k}\,[\,\bf{m}}\,P_{\bf{n}\,]}\,\delt \nonumber\\
{[}\,S_{\bf{m}\bf{n}}\,\xA,\,\Omega^{\bf{\rho}}\,\xB\,{]}&=&-\,i\,\frac{1}{2}\,(\,\gamma_{\bf{m}\bf{n}}\,)^{\bf{\rho}}_{\bf{\sigma}}\,\Omega^{\bf{\sigma}}\,\delt\nonumber\\
{[}\,S_{\bf{m}\bf{n}}\,\xA,\,\Sigma^{\bf{k}\bf{l}}\,\xB\,{]}&=
&-i\,\delta_{\bf mn}{}^{\bf kl}\,\deltP\,-\,i\delta_{[\,\bf{m}}{}^{[\,\bf{k}}\,\eta_{\bf{n}\,]\,\bf{s}}\Sigma^{\bf{l}]\bf{s}}\,\delt\nonumber\\
{\{}\,D_{\bf{\rho}}\,\xA,\,D_{\bf{\sigma}}\xB\,{\}}&=&i\,2\,(\,\gamma^{\bf{m}}\,)_{\bf{\rho\sigma}}\,P_{\bf{m}}\,\delt\nonumber\\
{[}\,D_{\bf{\rho}}\,\xA,\,P_{\bf{m}}\,\xB\,{]}&=&-\,i\,2\,(\,\gamma_{\bf{m}})_{\bf{\rho\sigma}}\,\Omega^{\bf{\sigma}}\,\delt\nonumber\\
{\{}\,D_{\bf{\rho}}\,\xA,\,\Omega^{\bf{\sigma}}\,\xB\,{\}}&=&-\,i\,\delta_{\bf{\rho}}^{\bf{\sigma}}\,\deltP\,-\,i\,\frac{1}{2}\,(\,\gamma_{\,\bf{m}\bf{n}\,}\,)_{\bf{\rho}}^{\bf{\sigma}}\,\Sigma^{\bf{mn}}\,\delt\nonumber\\
{[}\,D_{\bf{\rho}}\,\xA,\,\Sigma^{\bf{mn}}\,\xB\,{]}&=&0\nonumber\\
{[}\,P_{\bf{m}}\,\xA,\,P_{\bf{n}}\,\xB\,{]}&=
&-\,i\,\eta_{\bf mn}\,\deltP\,+\,i\,\eta_{\bf{m}\,[\bf{h}}\,\eta_{\bf{n}|\bf{s}]}\Sigma^{\bf hs}\,\delt\nonumber\\
{[}\,P_{\bf{m}}\,\xA,\,\Omega^{\bf{\rho}}\,\xB\,{]}&=&0\nonumber\\
{[}\,P_{\bf{m}}\,\xA,\,\Sigma^{\bf{k}\bf{l}}\,\xB\,{]}&=&0\nonumber\\
{\{}\,\Omega^{\bf \rho}\,\xA,\,\Omega^{\bf \sigma}\,\xB\,{\}}&=&0\nonumber\\
{[}\,\Omega^{\bf \rho}\,\xA,\,\Sigma^{\bf mn}\,\xB\,{]}&=&0\nonumber\\
{[}\,\Sigma^{\bf{m}\,\bf{n}}\,\xA,\,\Sigma^{\bf{k}\,\bf{l}}\,\xB\,{]}&=&0\nonumber\\
&{\xymatrix{\ar@/_/@{>}[r]&}}&\mbox{\footnotesize{Same for Left $\,\rightarrow\,$ Right}}\nonumber\\
{[}\,\mbox{Left},\,\mbox{Right}\,\}\,&=&\,0\nonumber
\end{eqnarray}
The only nonvanishing terms in the metric and structure constants are (as could be guessed by dimensional analysis)
\be
\eta_{PP} , \ \eta_{S\Sigma}, \ \eta_{D\Omega} , ; \quad f_{SPP} \ , \  f_{SS\Sigma} \ , \ f_{DDP}\ , \ f_{SD\Omega}
\ee
where we have lowered the upper index on $f$ with $\eta$ to take advantage of its total (graded)antisymmetry, and used ``schematic" notation, replacing explicit indices with their type:
\be
\label{schem}
\mbox{{\scriptsize\Fontamici{M}}}
\,\defeq\,(\,_{MN},\,_{\mu},\,_M,\,^{\mu},\,^{\,MN\,}\,)
\,\defeq\,(\,S,\,D,\,P,\,\Omega,\,\Sigma\,)
\ee
Explicitly these are, for the left-handed algebra,
\be
(\eta)_{\bf mn}\,=\,\eta_{\bf mn}\,,\, \  (\eta)_{\bf mn}{}^{\bf pq} \,=\, \delta_{\bf mn}{}^{\bf pq} \,,\,(\eta)_{\bf \sigma}{}^{\bf \rho}\,=\,\delta_{\bf \sigma}^{\bf \rho} \\
\ee
\be
f_{\bf mn}{}^{\bf p\,q}\,=\,-\,\delta_{\bf mn}{}^{\bf pq} \,,\, \ f_{\bf mn\,pq}{}^{\bf rs} \,=\, \eta_{\bf [m[p} \delta_{\bf q]n]}{}^{\bf rs} \,,\, f_{\bf \sigma\rho}{}^{\bf m}\,=\,2\,(\,\gamma^{\bf m}\,)_{\bf \sigma\rho}\,,\,f_{\bf mn\,\sigma}{}^{\bf \rho}\,=\,-\,\frac{1}{2}\,(\,\gamma_{\bf mn}\,)_{\bf \sigma}^{\bf \rho}
\ee
For the right-handed algebra we change the signs of the corresponding terms in $\eta_{\go{MN}}$ but not in $f$.

For dealing with antisymmetric pairs of indices we have introduced an implicit metric such that for any two antisymmetric tensors we have
\be
\label{dot}
A\cdot B \,\equiv\, \frac12 A^{\bf mn}B_{\bf mn}
\ee
The identity matrix with respect to this inner product is
\be
\delta_{\bf mn}{}^{\bf pq} \,\equiv\, \delta_{[\bf m}{}^{\bf p} \delta_{{\bf n}]}{}^{\bf q}
\ee
\subsection{Background fields}  
The aim is to find linearised formulation of the $3$D T-dual theory. We are following the approach used in the previous paper, see \cite{natural}, section $1$.$2$. We will briefly mention the outline here:

We want to use the T-dual formulation of the stringy generalisation of the oscillatory algebra (\ref{algebra}). We introduce the background fields via vielbeins. Following \cite{warren} but using algebra (\ref{twotwo}) we get:
\be
\Pi_{\mbox{{\scriptsize\Fontamici{A}}}}\xA\,=\,E_{\mbox{{\scriptsize\Fontamici{A}}}}{}^{{\mbox{{\scriptsize\Fontamici{M}}}}}(X^{\mbox{{\scriptsize\Fontamici{N}}}})Z_{\mbox{{\scriptsize\Fontamici{M}}}}
\ee
the affine Lie algebra for the $\Pi_{\mbox{{\scriptsize\Fontamici{A}}}}$ could be compactly written as:
\begin{equation}
\begin{array}{cccccc}
\label{Lenuska1}
\mbox{[}\Pi_{\mbox{{\scriptsize\Fontamici{A}}}}\xA,\Pi_{\mbox{{\scriptsize\Fontamici{C}}}}\xB\mbox{]}\,\equiv\,-i\eta_{\mbox{{\scriptsize\Fontamici{A}}}\mbox{{\scriptsize\Fontamici{C}}}}\,\deltP-iT_{\mbox{{\scriptsize\Fontamici{A}}}\mbox{{\scriptsize\Fontamici{C}}}}{}^{\mbox{{\scriptsize\Fontamici{E}}}}\Pi_{\mbox{{\scriptsize\Fontamici{E}}}}\,\delt
\end{array}
\end{equation}
where $T_{\mbox{{\scriptsize\Fontamici{A}}}\mbox{{\scriptsize\Fontamici{C}}}}{}^{\mbox{{\scriptsize\Fontamici{E}}}}$ is a (super)stringy generalisation of torsion, see \cite{natural}:
\be
\label{torsion}
T_{\mbox{{\scriptsize\Fontamici{A}}}\mbox{{\scriptsize\Fontamici{C}}}}{}^{\mbox{{\scriptsize\Fontamici{E}}}}=E_{[\mbox{{\scriptsize\Fontamici{A}}}}{}^{\mbox{{\scriptsize\Fontamici{M}}}}(D_{\mbox{{\scriptsize\Fontamici{M}}}}E_{\mbox{{\scriptsize\Fontamici{C}}})}{}^{\mbox{{\scriptsize\Fontamici{N}}}})E^{-1}_{\mbox{{\scriptsize\Fontamici{N}}}}{}^{\mbox{{\scriptsize\Fontamici{E}}}}
+\frac{1}{2}\eta^{\mbox{{\scriptsize\Fontamici{E}}}\mbox{{\scriptsize\Fontamici{D}}}}E_{\mbox{{\scriptsize\Fontamici{D}}}}{}^{\mbox{{\scriptsize\Fontamici{M}}}}(D_{\mbox{{\scriptsize\Fontamici{M}}}}E_{[\mbox{{\scriptsize\Fontamici{A}}}|}{}^{\mbox{{\scriptsize\Fontamici{N}}}})E^{-1}_{\mbox{{\scriptsize\Fontamici{N}}}}{}^{\mbox{{\scriptsize\Fontamici{F}}}}\eta_{\mbox{{\scriptsize\Fontamici{F}}}|\mbox{{\scriptsize\Fontamici{C}}})}
+E_{\mbox{{\scriptsize\Fontamici{A}}}}{}^{\mbox{{\scriptsize\Fontamici{M}}}}E_{\mbox{{\scriptsize\Fontamici{C}}}}{}^{\mbox{{\scriptsize\Fontamici{N}}}}E^{-1}_{\mbox{{\scriptsize\Fontamici{P}}}}{}^{\mbox{{\scriptsize\Fontamici{E}}}}f_{\mbox{{\scriptsize\Fontamici{M}}}\mbox{{\scriptsize\Fontamici{N}}}}{}^{\mbox{{\scriptsize\Fontamici{P}}}}
\ee
where $[\,\go A\,|\,|\,\go C\,)$ indicates graded antisymmetrization in only those indices. By the $\Ds_{\Ms}$ in the (\ref{torsion}) and in the whole text we mean the group covariant derivatives of the (non-affine) part of algebra (\ref{algebra}): $[\,\Ds_{\Ms},\,\Ds_{\Ns}\,\}\,=\,-\,i\,f_{\Ms\,\Ns}{}^{\Es}\,\Ds_{\Es}$.
 
Note that the (super)Jacobi identities imply the total graded antisymmetry of the torsion, just as for the structure constants. Torsion (\ref{torsion}) can be identified with that of ``ordinary" curved-space covariant derivatives by use of the strong constraint, as explained in \cite{natural}, \cite{warren}. 

We can set the coefficient of the Schwinger term to be the metric $\eta$, the vielbein is forced to obey the orthogonality constraints: 
\begin{equation}
\label{equation}
E_{\mbox{{\scriptsize\Fontamici{A}}}}{}^{\mbox{{\scriptsize\Fontamici{M}}}}\eta_{\mbox{{\scriptsize\Fontamici{M}}}\mbox{{\scriptsize\Fontamici{N}}}}\,E_{\,\mbox{{\scriptsize\Fontamici{C}}}}{}^{\mbox{{\scriptsize\Fontamici{N}}}}\,\equiv\,\eta_{\mbox{{\scriptsize\Fontamici{A}}}\mbox{{\scriptsize\Fontamici{C}}}}
\end{equation}
This choice  does not affect the physics, and simplifies many of the expressions. For example, it implies the total graded antisymmetry of the torsion, when the upper index is implicitly lowered with $\eta$:
\be
\label{t2}
T_{\go{A\,B\,C}} \,=\,\frac12 E_{[\,\go A\,|}{}^{\go M}(D_{\go M} E_{|\,\go B}{}^{\go N})E_{\go C\,)\,\go N}
	+ E_{\go A}{}^{\go M} E_{\go B}{}^{\go N} E_{\go C}{}^{\go P} f_{\go{M\,N\,P}}
\ee
where we have used $E^{-1}_{\,\,\,\mbox{{\scriptsize\Fontamici{M}}}}{}^{\,\mbox{{\scriptsize\Fontamici{A}}}}\,=\,\eta^{\mbox{{\scriptsize\Fontamici{A\ B}}}}\eta_{\mbox{{\scriptsize\Fontamici{M\ N}}}}E_{\,\mbox{{\scriptsize\Fontamici{B}}}}{}^{\mbox{{\scriptsize\Fontamici{N}}}}$.  (Also note that in the first term the graded antisymmetrization can be written as a cyclic sum without the $1/2$, since it is already graded antisymmetric in the last two indices.)  Thus, because of orthogonality, the vielbein is like (the exponential of) a super $2$-form, while the torsion is a super $3$-form; similarly, the Bianchi identities are a super $4$-form.

The (super)orthogonality constraint (\ref{equation}) could be fully solved for the general structure of the vielbein $E_{\go A}{}^{\go M}$. However, we are interested just in the linear level. Thus we get the (super)orthogonality constraint for the linearised part of the vielbein $E^{\scriptstyle (1)}{}_{\go A}{}^{\go M}$:
\ba
E_{\go A}{}^{\go M}&=&\delta_{\go A}{}^{\go M}\,+\,E^{\scriptstyle (1)}{}_{\go A}{}^{\go M}\,+\,\Oo\,(\,E^{\scriptstyle (2)}\,)\,\Rightarrow\,\text{using (\ref{equation})}\\
E^{\scriptstyle (1)}{}_{(\go{\,A\,B\,}]}&=&0
\ea
We would also need the linear level version of the equation (\ref{t2}):
\ba
T_{\go {A\,B\,C}}&=&f_{\go {A\,B\,C}}\,+\,T^{\scriptstyle (1)}{}_{\go {A\,B\,C}}\,+\,\Oo\,(\,E^{\scriptstyle (2)}\,)\\
\label{T1}
\mbox{where}\,\,\,\,T^{{\scriptstyle (1)}}{}_{\As\,\Bs\,\Cs}&\defeq&\frac{1}{2}\,D_{[\,\As}\,E^{{\scriptstyle (1)}}{}_{\Bs\,\Cs\,)}\,+\,\frac{1}{2}\,\Eone{}_{[\As}{}^{\Ms}\,f_{\Ms\,|\Bs\,\Cs\,)}
\ea
\subsection{Further constraints and gauge fixing}
Following the discussion in the subsection $4$.$2$ in the paper \cite{natural}, we get the coset constraint on the torsion piece $T_{\go {S\,A}}{}^{\go B}\,=\,f_{\go {S\,A}}{}^{\go B}$ (where we used the ${\go S}$ index as the schematic index (\ref{schem}) and ${\go {A,\,B}}$ are general indices). On the linear level the previous condition becomes: $T^{\scriptstyle{(1)}}{}_{\go {S\,A\,B}}\,=\,0$. From this one gets the condition for the linear vielbein: $E^{\scriptstyle (1)}{}_{\go S}{}^{\go M}\,=\,0\,+\,\Oo\,(\,E^{\scriptstyle{(2)}}\,)$.

We would like to gauge fix some of the remaining gauge freedom. Note that the coset constraints discussed above sets the gauge parameter (defined below) $\lambda_{\go{S}}\,=\,0$. From specific gauge fixing we get the further conditions on the linear vielbein $E^{\scriptstyle{(1)}}$. The gauge transformations are given as (see also \cite{warren}):
\ba
\delta_{\Lambda}\,\Pi_{\go A}\,=\,[\,-\,i\,\Lambda,\,\Pi_{\go A}\,\}\\
\mbox{where}\,\,\Lambda\,\defeq\,\int\,d1\,\lambda^{\go M}\,(\,X\,)\,D_{\go M}\,\nonumber
\ea
We are working in the basis where the covariant derivatives satisfy: $[\,{D}_{\go M},\,{D}_{\go N}\,\}\,=\,i\,f_{\go{M\,N}}{}^{\go P}\,{D}_{\go P}$. Thus the (linear)gauge transformation of (linear)vielbein are: 
\be
\label{nunu}
\delta_{\Lambda}\,E^{\scriptstyle{(1)}}{}_{\go{A\,B}}\,=\,-\,\frac{i}{2}\,D_{[\,\go{A}}\,\lambda_{\go{B}\,)}\,+\,f_{\go{A\,B}}{}^{\go{C}}\,\lambda_{\go{C}}
\ee
Now, we can pick the following gauge:
\ba
\label{gf1}\gamma_{\bf a}{}^{\bf{\alpha\,\beta}}\,E^{\scriptstyle{(1)}}{}_{\bf{\alpha\,\beta}}\,=\,0\,&\Rightarrow&\,\lambda_{\bf{a}}\,\propto\,\gamma_{\bf{a}}{}^{\bf{\alpha\,\beta}}\,D_{\bf{\alpha}}\,\lambda_{\bf{\beta}}\\
\label{gf2}\gamma^{\bf{a\,\,\alpha\,\beta}}\,E^{\scriptstyle{(1)}}{}_{\bf{\alpha\,a}}\,=\,0\,&\Rightarrow&\,\lambda^{\bf{\alpha}}\,\propto\,\gamma^{\bf{a\,\,\alpha\,\beta}}\,D_{[\,\bf{a}}\,\lambda_{\bf{\beta}\,]}\\
\label{gf3}E^{\scriptstyle{(1)}}{}_{\bf{a\,b}}\,=\,0\,&\Rightarrow&\,\lambda_{\bf{a\,b}}\,\propto\,D_{[\,\bf{a}}\,\lambda_{\bf{b}\,]}\\
&&{\scriptstyle {\xymatrix{\ar@/_/@{>}[r]&}} \mbox{\footnotesize{Same for Left $\,\rightarrow\,$ Right}}}\nonumber
\ea
We can see that by (\ref{gf1}), (\ref{gf2}), (\ref{gf3}) we automatically have expressions for gauge parameters $\lambda_{P},\,\lambda_{\Omega},\,\lambda_{\Sigma}$ as derivatives of another gauge parameter $\lambda_{D}$. It is unlike the usual $\sN\,=\,1$ supergravity where we need first to solve the chirality condition to relate derivatives of $\Lambda$ with $K$, see section X.A.$1$ in \cite{fields}, also section $5$.$3$ in \cite{1001}. Moreover, the (\ref{gf1})-(\ref{gf3}) give the constraints on $E^{\scriptstyle{(1)}}$ and solving those we will get:
\ba
E^{\scriptstyle{(1)}}{}_{DD}\,=\,E^{\scriptstyle{(1)}}{}_{\alpha\,\beta}\,=\,0,\,\,\,\,E^{\scriptstyle{(1)}}{}_{PP}\,=\,E^{\scriptstyle{(1)}}{}_{\bf{a\,b}}\,=0,\,\,\,\,E^{\scriptstyle{(1)}}{}^{\bf{\alpha\,\beta}}{}_{\bf{\beta}}\,=\,0\,\,\mbox{(part of $E^{\scriptstyle{(1)}}{}^{P}{}_{D}$)}
\ea 
Later (by dimension $-\frac{1}{2}$ constraints) one can see that the $E^{\scriptstyle{(1)}}{}_{PD}\,=\,0$. We thus need to set up the dimensional constraints. The following table {\texttt{(\ref{hrncek})}} summarise the torsion dimensions:

\begin{table}[H]
{\centering
\begin{tabular}{| cc |}
\noalign{\vspace{.1in}}
\hline
{ \texttt{Torsion}}&{ \texttt{Dim.}}\\
\hline
 ${\color{red} T_{S\,S}^{\,\,\,\,\,\,\,\,\,\Sigma}}$&${\color{red}-\,2}$\\
\hline
${\color{red} T_{S\,S}{}^{\Omega}}$&${\color{red} -\,\frac{3}{2}}$\\
\hline
${\color{red} T_{S\,S}^{\,\,\,\,\,\,\,\,\,P}}$&${\color{red} -\,1}$\\
\hline
${\color{red} T_{S\,D}{}^{\Omega}}$&${\color{red} -\,1}$\\
\hline
${\color{red} T_{S\,S}{}^{D}}$&${\color{red} -\,\frac{1}{2}}$\\
\hline
${\color{red} T_{S\,P}{}^{\Omega}}$&${\color{red} -\,\frac{1}{2}}$\\
\hline
${\color{red} T_{D\,D}{}^{\Omega}}$&${\color{red} -\,\frac{1}{2}}$\\
\hline
${\color{blue} T_{S\,S}^{\,\,\,\,\,\,\,\,\,S}}$&${\color{blue} 0}$\\
\hline
${\color{blue} T_{S\,D}{}^{D}}$&${\color{blue} 0}$\\
\hline
${\color{blue} T_{S\,P}^{\,\,\,\,\,\,\,\,\,P}}$&${\color{blue} 0}$\\
\hline
${\color{blue} T_{D\,D}{}^{P}}$&${\color{blue} 0}$\\
\hline
\noalign{\vspace{.1in}}
\end{tabular}
\quad
\begin{tabular}{| cc |}
\noalign{\vspace{.1in}}
\hline
{ \texttt{Torsion}}&{ \texttt{Dim.}}\\
\hline
${\color{OliveGreen} T_{S\,D}{}^{S}}$&${\color{OliveGreen} \frac{1}{2}}$\\
\hline
${\color{OliveGreen} T_{S\,P}{}^{D}}$&${\color{OliveGreen} \frac{1}{2}}$\\
\hline
${\color{OliveGreen} T_{D\,D}{}^{D}}$&${\color{OliveGreen} \frac{1}{2}}$\\
\hline
${\color{OliveGreen} T_{P\,P}{}^{\Omega}}$&${\color{OliveGreen}\frac{1}{2}}$\\
\hline
${\color{Magenta} T_{S\,P}^{\,\,\,\,\,\,\,\,\,S}}$&${\color{Magenta}\scriptstyle1}$\\
\hline
${\color{Magenta} T_{S\,\Omega}{}^{D}}$&${\color{Magenta}1}$\\
\hline
${\color{Magenta} T_{D\,P}{}^{D}}$&${\color{Magenta}1}$\\
\hline
${\color{Magenta} T_{D\,D}{}^{S}}$&${\color{Magenta}1}$\\
\hline
${\color{Magenta} T_{P\,P}^{\,\,\,\,\,\,\,\,\,P}}$&${\color{Magenta}1}$\\
\hline
${ T_{S\,\Omega}{}^{S}}$&${\frac{3}{2}}$\\
\hline
${ T_{D\,P}{}^{S}}$&${\frac{3}{2}}$\\
\hline
${ T_{D\,\Omega}{}^{D}}$&${\frac{3}{2}}$\\
\hline
${ T_{P\,P}{}^{D}}$&${\frac{3}{2}}$\\
\hline
\end{tabular}
\quad
\begin{tabular}{| cc |}
\noalign{\vspace{.1in}}
\hline
{ \texttt{Torsion}}&{ \texttt{Dim.}}\\
\hline
${ T_{S\,\Sigma}^{\,\,\,\,\,\,\,\,\,S}}$&${2}$\\
\hline
${ T_{D\,\Omega}{}^{S}}$&${2}$\\
\hline
${ T_{P\,P}^{\,\,\,\,\,\,\,\,\,S}}$&${2}$\\
\hline
${ T_{P\,\Omega}{}^{D}}$&${2}$\\
\hline
${ T_{D\,\Sigma}{}^{S}}$&${\frac{5}{2}}$\\
\hline
${T_{P\,\Omega}{}^{S}}$&${\frac{5}{2}}$\\
\hline
${ T_{\Omega\,\Omega}{}^{D}}$&${\frac{5}{2}}$\\
\hline
${ T_{P\,\Sigma}^{\,\,\,\,\,\,\,\,\,S}}$&${3}$\\
\hline
${ T_{\Omega\,\Omega}{}^{S}}$&${3}$\\
\hline
${T_{\Omega\,\Sigma}{}^{S}}$&${\frac{7}{2}}$\\
\hline
${ T_{\Sigma\,\Sigma}^{\,\,\,\,\,\,\,\,\,S}}$&${4}$\\
\hline
\noalign{\vspace{.1in}}
\end{tabular} 
\caption{Torsion dimensions \protect\label{hrncek}}
}
\end{table}

Notice that many of the torsions in the previous table \texttt{(\ref{hrncek})} are fixed (to flat structure constants $f_{\go {S\,A}}{}^{\go B}$). 

We put the torsions of the negative (engineering) dimensions to $0$ (as always in QFT, see the {\color{red} red} coloured torsions in the previous table). We also put the (unfixed) torsions of the zero dimension to $0$ (see the {\color{blue} blue} torsions in the previous table), see \cite{natural}. We will also put the dimension $\frac{1}{2}$ (unfixed) torsions to $0$ (the {\color{OliveGreen} green} torsions in the table). Doing that we produce just algebraic constraints on veilbeins.

The nontrivial dimensional constraints are:
\be
{{\color{red}T_{D\,D}{}^{\Omega}}\,=\,0},\,\,{\color{blue}T_{D\,D}{}^{P}}\,=\,f_{D\,D}{}^{P},\,\,{\color{OliveGreen}T_{D\,D}{}^{D}}\,=\,0,\,\,{\color{OliveGreen}T_{P\,P}{}^{\Omega}}\,=\,0
\ee
\subsection{Dimensional constraints: solution}
The solution of the previous nontrivial dimensional constraints could be given in a full generality, however in this paper we are interested just in the linearised case. The tables {\texttt{(\ref{hrncek1})}} and {\texttt{(\ref{hrncek2})}} summarise the linearised solutions of those four constraints (notice that we have also the possibility of mixed left/right indices):
\begin{table}[H]
{\centering
\begin{tabular}{| ccc |}
\noalign{\vspace{.1in}}
\hline
${{{\color{red}T_{D\,D}{}^{\Omega}}\,=\,0}}\,\,\,\mbox{and}\,\,\,\gamma^{\bf{a\,\,\alpha\,\beta}}\,E^{\scriptstyle{(1)}}{}_{\bf{\alpha\,a}}\,=\,0$&$\Rightarrow$&$E^{\scriptstyle{(1)}}{}_{PD}\,=\,0$\\
\hline
${\color{blue}T_{D\,D}{}^{P}}\,=\,f_{D\,D}{}^{P}$&$\Rightarrow$&$E^{\scriptstyle{(1)}}{}_{D\Omega}\,=\,0$\\
\hline
${\color{OliveGreen}T_{P\,P}{}^{\Omega}}\,=\,0\,\,\,\mbox{or}\,\,\,{\color{OliveGreen}T_{D\,D}{}^{D}}\,=\,0$&$\Rightarrow$&$E^{\scriptstyle{(1)}}{}_{\Sigma\,D}\,=\,E^{\scriptstyle{(1)}}{}^{\bf{ab}}{}_{\bf{\alpha}}\,=\,-\,2\,\gamma^{\bf{[\,a}}{}_{\bf{\alpha\,\rho}}\,E^{\scriptstyle{(1)}}{}^{\bf{\rho\,|\,b\,]}}\,$\\
&&$\equiv\,\gamma\,\rp\,E^{\scriptstyle{(1)}}{}_{\Omega}{}^{P}$\\
\hline
\end{tabular}
}
\caption{Unmixed constraints}
\protect\label{hrncek1}
\end{table}
\begin{table}[H]
{\centering
\begin{tabular}{| ccc |}
\noalign{\vspace{.1in}}
\hline
${{{\color{red}T_{D\,D}{}^{\tilde{\Omega}}}\,=\,0}}$&$\Rightarrow$&$E^{\scriptstyle{(1)}}{}_{P\,\tilde{D}}\,\equiv\,E^{\scriptstyle{(1)}}{}_{\bf{a\,\tilde{\alpha}}}\,=\,-\,\frac{1}{2}\,\gamma_{\bf{a}}{}^{\bf{\beta\,\epsilon}}\,D_{\bf{\beta}}\,E^{\scriptstyle{(1)}}{}_{\bf{\epsilon\,\tilde{\alpha}}}\,\equiv\,-\,\gamma\,{\color{red} . }\,D_{D}\,{\color{red} . }\,E^{\scriptstyle{(1)}}{}_{D\,\tilde{D}}$\\
\hline
${\color{blue}T_{D\,D}{}^{\tilde{P}}}\,=\,0$&$\Rightarrow$&$E^{\scriptstyle{(1)}}{}_{P\,\tilde{P}}\,\equiv\,E^{\scriptstyle{(1)}}{}_{\bf{a\,\tilde{b}}}\,=\,-\,\frac{1}{2}\,\gamma_{\bf{a}}{}^{\bf{\beta\,\epsilon}}\,D_{\bf{\beta}}\,E^{\scriptstyle{(1)}}{}_{\bf{\epsilon\,\tilde{b}}}\,\equiv\,-\,\gamma\,\rp\,D_{D}\,\rp\,E^{\scriptstyle{(1)}}{}_{D\,\tilde{P}}$\\
\hline
${\color{blue}T_{D\,\tilde{D}}{}^{P}}\,=\,0$&$\Rightarrow$&$E^{\scriptstyle{(1)}}{}_{\Omega\,\tilde{D}}\,\equiv\,E^{\scriptstyle{(1)}}{}^{\bf{\alpha}}{}_{\bf{\tilde{\beta}}}\,=\,-\,\frac{1}{6}\,\gamma^{\bf{a}\,}{}^{\bf{\epsilon\,\alpha}}\,D_{\bf{[\epsilon}}\,E^{\scriptstyle{(1)}}{}_{\bf{a]\,\tilde{\beta}}}\,\equiv\,-\,\gamma\,\rp\,D_{[D}\,\rp\,E^{\scriptstyle{(1)}}{}_{P]\,\tilde{D}}$\\
\hline
${\color{OliveGreen}T_{P\,\tilde{P}}{}^{\Omega}}\,=\,0$&$\Rightarrow$&$E^{\scriptstyle{(1)}}{}_{\Omega\,\tilde{P}}\,\equiv\,E^{\scriptstyle{(1)}}{}^{\bf{\alpha}}{}_{\bf{\tilde{a}}}\,=\,-\,\frac{1}{6}\,\gamma^{\bf{b}\,}{}^{\bf{\epsilon\,\alpha}}\,D_{\bf{[\epsilon}}\,E^{\scriptstyle{(1)}}{}_{\bf{b]\,\tilde{a}}}\,\equiv\,-\,\gamma\,\rp\,D_{[D}\,\rp\,E^{\scriptstyle{(1)}}{}_{P]\,\tilde{P}}$\\
\hline
${\color{OliveGreen}T_{P\,P}{}^{\tilde{\Omega}}}\,=\,0$&$\Rightarrow$&$E^{\scriptstyle{(1)}}{}_{\Sigma\,\tilde{D}}\,\equiv\,E^{\scriptstyle{(1)}}{}^{\bf{ab}}{}_{\bf{\tilde{\alpha}}}\,=\,\eta^{\bf{a\,c}}\,\eta^{\bf{b\,d}}\,D_{\bf{[c}}\,E^{\scriptstyle{(1)}}{}_{\bf{d]\,\tilde{\alpha}}}\,\equiv\,\eta\,\eta\,\rp\,D_{[P}\,\rp\,E^{\scriptstyle{(1)}}{}_{P]\,\tilde{D}}$\\
\hline
\end{tabular}
}
\caption{Mixed constraints}
\protect\label{hrncek2}
\end{table}
(Notice: the ``$\rp$'' in the previous tables means the symbolic contraction. Do not be confused with ``$\cdot$'' defined in (\ref{dot}).)

From the table {\color{black}\texttt{(\ref{hrncek1})}} we can see that we have one linear relation: $E^{\scriptstyle{(1)}}{}_{\Sigma\,D}\,\propto\,E^{\scriptstyle{(1)}}{}_{\Omega\,P}$. From the table {\color{black}\texttt{(\ref{hrncek2})}} we have linear relations: $\{\,E^{\scriptstyle{(1)}}{}_{P\,\tilde{D}},\,E^{\scriptstyle{(1)}}{}_{P\,\tilde{P}},\,E^{\scriptstyle{(1)}}{}_{\Omega\,\tilde{D}},\,E^{\scriptstyle{(1)}}{}_{\Omega\,\tilde{P}},\,E^{\scriptstyle{(1)}}{}_{\Sigma\,\tilde{D}}\,\}\,\propto\,E^{\scriptstyle{(1)}}{}_{D\,\tilde{D}}$. Again, we have automatically obtained the expressions for the vielbeins as derivatives of $\Eone{}_{D\,\tilde{D}}$ vielbein (pre potential). It is unlike the $\sN\,=\,1$ supergravity where the pre potential comes as the solution of the bisection condition (or chirality condition in covariant approach), see section X.A.$1$ in \cite{fields} and section $5$.$2$.a and $5$.$3$ in \cite{1001}. 
\subsection{\texorpdfstring{\textcolor{Magenta}{{Dimension $1$}}}{{Dimension 1}} unmixed constraints}
To proceed we need to find the constraints for the dimension $1$ torsions. We can see that putting those to zero in general introduces the differential constraints, that we do not want (except of the strong constraint and later the equation of motion). However there is a way how to fix dimension $1$ torsions without producing differential constraints. We will use the following set of unmixed constraints (again we have two cases for the torsion index structure: mixed and unmixed):
\ba
\label{unmx1}
{\color{Magenta}T^{\scriptstyle{(1)}}{}_{P\,P\,P}}&\equiv&T^{\scriptstyle{(1)}}{}_{\bf{a\,b\,c}}\,=\,\vartheta\,\varepsilon_{\bf{a\,b\,c}}\,B\\
{\color{Magenta}T^{\scriptstyle{(1)}}{}_{D\,D\,\Sigma}}&\equiv&T^{\scriptstyle{(1)}}{}_{\bf{\alpha\,\beta}}{}^{\bf{ab}}\,=\,\xi\,\gamma^{\bf{ab}}{}_{\bf{\alpha\,\beta}}\,B\,\nonumber\\
{\color{Magenta}T^{\scriptstyle{(1)}}{}_{P\,D\,\Omega}}&\equiv&T^{\scriptstyle{(1)}}{}_{\bf{a\,\alpha}}{}^{\bf{\beta}}\,=\,\zeta\,\gamma_{\bf{a}\,}{}_{\bf{\alpha}}^{\bf{\beta}}\,B\nonumber
\ea
where the new object $B$ is determined from (\ref{unmx1}). Using the linearised Bianchi identity we get:
\ba
\label{linB}
&\mbox{\texttt{ lin. Bianchi id.}}&D_{[\As}\,T^{\scriptstyle{(1)}}{}_{\Bs\,\Cs)}{}^{\Ds}\,-\,f_{[\As\,\Bs|}{}^{\Ms}\,T^{\scriptstyle{(1)}}{}_{\Ms|\,\Cs)}{}^{\Ds}\,-\,T^{\scriptstyle{(1)}}{}_{[\As\,\Bs}{}^{\Ms}\,f_{\Ms|\,\Cs)}{}^{\Ds}\,=\,0\,\nonumber\\
&\mbox{\texttt{for dim. $1$:}}\,\,&\frac{1}{2}\,T^{\scriptstyle{(1)}}{}_{\bf{\alpha\,\beta}}{}^{\bf{ab}}\,-\,\gamma^{[\bf{a}}{}_{(\bf{\alpha}|\,\epsilon}\,T^{\scriptstyle{(1)}}{}_{\bf{\beta})}{}^{\bf{\epsilon\,|b}]}\,+\,\gamma^{\bf{c}}{}_{\bf{\alpha\,\beta}}\,T^{\scriptstyle{(1)}}{}_{\bf{c}}{}^{\bf{a\,b}}\,=\,0\,\,\nonumber\\
&\mbox{\texttt{using} (\ref{unmx1}){\texttt{:}}}\,\,&\frac{\xi}{2}\,\gamma^{\bf{ab}}{}_{\bf{\alpha\,\beta}}\,+\,\zeta\,\gamma^{[\bf{a}|\,}{}_{(\bf{\alpha}}^{\,\,\bf{\epsilon}}\,\gamma^{\bf{b}]\,}{}_{\bf{\beta})\,\bf{\epsilon}}\,+\,\vartheta\,\varepsilon^{\bf{a\,b\,c}}\,\gamma_{\bf{c}\,\,\bf{\alpha\,\beta}}\,=\,0\,\,\Rightarrow\nonumber\\
&&\xi\,=\,-\,8\,\zeta\,-\,2\,\vartheta\\
\nonumber
\ea
Equation (\ref{T1}) gives explicit relations for the $T^{\scriptstyle{(1)}}$'s from (\ref{unmx1}): 
\ba
\label{hu}
T^{\scriptstyle{(1)}}{}_{\bf{a\,b\,c}}&=&-\,\frac{1}{2}\,\eta_{\bf{[a|\,d}}\,\eta_{\bf{b}|\,e}\,E^{\scriptstyle{(1)}}{}^{\bf{de}}{}_{\bf{c]}}\,\equiv\,-\,\eta\,\eta\,\rp\,E^{\scriptstyle{(1)}}{}_{\Sigma\,P}\\
\label{huhu}
T^{\scriptstyle{(1)}}{}_{\bf{\alpha\,\beta}}{}^{\bf{ab}}&=&D_{(\bf{\alpha}}\,E^{\scriptstyle{(1)}}{}_{\bf{\beta})}{}^{\bf{ab}}\,+\,2\,\gamma^{\bf{c}}{}_{\bf{\alpha\,\beta}}\,E^{\scriptstyle{(1)}}{}^{\bf{ab}}{}_{\bf{c}}\,\equiv\,D\rp\,E^{\scriptstyle{(1)}}{}_{D\,\Sigma}\,+\,\gamma\rp\,E^{\scriptstyle{(1)}}{}_{\Sigma\,P}\\
\label{huhuhu}
T^{\scriptstyle{(1)}}{}_{\bf{a\,\alpha}}{}^{\bf{\beta}}&=&D_{\bf{\alpha}}\,E^{\scriptstyle{(1)}}{}^{\,\bf{\beta}}{}_{\bf{a}}\,+\,2\,\gamma_{\bf{a\,\,\alpha\,\epsilon}}\,E^{\scriptstyle{(1)}}{}^{\,\,\bf{\epsilon\,\beta}}\,+\,\frac{1}{4}\,\varepsilon_{\bf{c\,d\,e}}\,\gamma^{\bf{e\,\,}}{}_{\bf{\alpha}}^{\bf{\beta}}\,E^{\scriptstyle{(1)}}{}^{\,\,\bf{cd}}{}_{\bf{a}}\\
&\equiv&D\rp\,E^{\scriptstyle{(1)}}{}_{\Omega\,P}\,+\,\gamma\rp\,E^{\scriptstyle{(1)}}{}_{\Omega\,\Omega}\,+\,\varepsilon\,\gamma\rp\,E^{\scriptstyle{(1)}}{}_{\Sigma\,P}\nonumber
\ea
From (\ref{hu}) and first relation of (\ref{unmx1}) we get: 
\be
\label{uhorka}
2\,\vartheta\,B\,=\,-\,\varepsilon_{\bf{h\,d\,e}}\,\eta^{\bf{h\,c}}\,E^{\scriptstyle{(1)}}{}^{\bf{de}}{}_{\bf{c}}
\ee
From (\ref{huhuhu}) and requiring that we want just algebraic constraints we get the second equation for (\ref{unmx1}) fixing constants: 
\be
\label{fazulka}
0\,=\,-\,3\,\zeta\,-\,\frac{1}{8}\,(\,2\,\vartheta\,+\,3\,\xi\,)\,-\,\frac{1}{2}\,\vartheta
\ee
Substituing result (\ref{linB}) we have soultion for any $\vartheta$ and $\zeta$ except when $\vartheta\,=\,-\,6\,\zeta$. That condition would produce the differential constraint on $\Eone{}_{\Omega\,P}$ (see eq. (\ref{Milano})). From (\ref{huhuhu}) and third of (\ref{unmx1}) we will get fixing of $E^{\scriptstyle{(1)}}{}_{\Omega\,\Omega}$. From (\ref{huhu}) and second of (\ref{unmx1}) we will get fixing of $E^{\scriptstyle{(1)}}{}_{P\,\Sigma}$. The net result of {dimension $1$} unmixed algebraic constraints (\ref{unmx1}) is that everything could be expressed in terms of $E^{\scriptstyle{(1)}}{}_{D\,\Sigma}$ and so (see table {\color{black}\texttt{(\ref{hrncek1})}}) by $E^{\scriptstyle{(1)}}{}_{P\,\Omega}$ (and two constants $\vartheta,\,\zeta$ s.t. $\vartheta\,\neq\,-\,6\,\zeta$):
\ba
\label{fraktura}
B&=&-\,\frac{1}{\vartheta\,+\,6\,\zeta}\,\gamma^{\bf{a}\,\,}{}^{\bf{\alpha\,}}_{\bf{\beta}}\,D_{\bf{\alpha}}\,E^{\scriptstyle{(1)}}{}^{\bf{\beta}}{}_{\bf{a}}\label{Milano}\\
&\equiv&\,-\,\gamma\,D_{D}\rp\,E^{\scriptstyle{(1)}}{}_{\Omega\,P}\nonumber\\
E^{\scriptstyle{(1)}}{}_{\Omega\,\Omega}\,=\,E^{\scriptstyle{(1)}}{}^{\,\bf{\alpha\,\beta}}&=&\frac{1}{12}\,\gamma^{\bf{a}\,\,(\alpha|\,\epsilon}\,D_{\bf{\epsilon}}\,E^{\scriptstyle{(1)}}{}^{\,\bf{\beta})}{}_{\bf{a}}\,+\,\frac{1}{12}\,\gamma_{\bf{a}}{}^{\bf{\alpha\,\beta}}E^{\scriptstyle{(1)}}{}^{\bf{ab}}{}_{\bf{b}}\\
&\equiv&\gamma\,D_{D}\rp\,E^{\scriptstyle{(1)}}{}_{\Omega\,P}\,+\,\gamma\rp\,E^{\scriptstyle{(1)}}{}_{\Sigma\,P}\nonumber\\
E^{\scriptstyle{(1)}}{}_{P\,\Sigma}\,=\,E^{\scriptstyle{(1)}}{}_{\bf{c}}{}^{\bf{ab}}&=&\,-\,\frac{1}{2}\,\gamma_{\bf{c}}{}^{\bf{\alpha\,\beta}}\,D_{\bf{\alpha}}\,E^{\scriptstyle{(1)}}{}_{\bf{\beta}}{}^{\bf{ab}}\,+\,(\,\vartheta\,+\,4\,\zeta\,)\,\eta_{\bf{c\,e}}\,\varepsilon^{\bf{e\,a\,b}}\,B\\
&\equiv&-\,\gamma\,D_{D}\rp\,E^{\scriptstyle{(1)}}{}_{D\,\Sigma}\,+\,\eta\,\varepsilon\rp\,B\nonumber 
\ea
\subsection{\texorpdfstring{\textcolor{Magenta}{{Dimension $1$}}}{{Dimension 1}} mixed constraints}
Some of the mixed dimension $1$ torsions are determined in terms of $E^{\scriptstyle{(1)}}{}_{\bf{\alpha}\,\bf{\tilde{\beta}}}\,\equiv\,E^{\scriptstyle{(1)}}{}_{D\,\tilde{D}}$ and $E^{\scriptstyle{(1)}}{}_{P\,\Omega}$ already. Using the previous results (tables {\color{black}\texttt{(\ref{hrncek1})}}, {\color{black}\texttt{(\ref{hrncek2})}} and results of previous section) we can see that mixed dimension $1$ torsions $T^{\scriptstyle{(1)}}{}_{\bf{a}\,\bf{\tilde{\alpha}}}{}^{\bf{\rho}}\,\equiv\,T^{\scriptstyle{(1)}}{}_{P\,\tilde{D}\,\Omega}$ and $T^{\scriptstyle{(1)}}{}_{\bf{\tilde{\alpha}}\,\beta}{}^{\bf{ab}}\,\equiv\,T^{\scriptstyle{(1)}}{}_{\tilde{D}\,D\,\Sigma}$ are fully determined, see (\ref{mx}). The mixed determined and undetermined torsions are summarised below: 
\ba
\begin{rcases}
\label{mx}
{\color{Magenta}T^{\scriptstyle{(1)}}{}_{P\,\tilde{D}\,\Omega}}\equiv T^{\scriptstyle{(1)}}{}_{\bf{a}\,\tilde{\bf{\alpha}}}{}^{\bf{\beta}}&=&D_{\bf{a}}\,E^{\scriptstyle{(1)}}{}_{\tilde{\bf{\alpha}}}{}^{\bf{\beta}}\,+\,D^{(\bf{\beta}}\,E^{\scriptstyle{(1)}}{}_{\tilde{\bf{\alpha}})\,\bf{a}}\equiv\,D_{P}\,E^{\scriptstyle{(1)}}{}_{\tilde{D}\,\Omega}\,+\,D_{(\Omega}\,E^{\scriptstyle{(1)}}{}_{\tilde{D})P}\\
{\color{Magenta}T^{\scriptstyle{(1)}}{}_{\tilde{D}\,D\,\Sigma}}\equiv T^{\scriptstyle{(1)}}{}_{\tilde{\bf{\alpha}}\,\beta}{}^{\bf{ab}}&=&D_{(\tilde{\bf{\alpha}}}\,E^{\scriptstyle{(1)}}{}_{\bf{\beta})}{}^{\bf{ab}}\,+\,D^{\bf{ab}}\,E^{\scriptstyle{(1)}}{}_{\tilde{\bf{\alpha}}\,\bf{\beta}}\equiv\,D_{(\tilde{D}}\,E^{\scriptstyle{(1)}}{}_{D)\Sigma}\,+\,D_{\Sigma}\,E^{\scriptstyle{(1)}}{}_{\tilde{D}D}\\
\end{rcases}
\ea
\ba
\begin{rcases}
\label{unmx}
{\color{Magenta}T^{\scriptstyle{(1)}}{}_{\tilde{P}\,P\,P}}\equiv\,T^{\scriptstyle{(1)}}{}_{\tilde{\bf{a}}\,\bf{b}\,\bf{c}}&=&D_{[\bf{b}}\,E^{\scriptstyle{(1)}}{}_{{\bf{c}}]\,\tilde{\bf{a}}}\,-\,\eta_{\bf{b}\,\bf{d}}\,\eta_{\bf{c}\,\bf{e}}\,E^{\scriptstyle{(1)}}{}^{\bf{de}}{}_{\tilde{\bf{a}}}\equiv\,D_{[P}\,E^{\scriptstyle{(1)}}{}_{P]\tilde{P}}\,-\,\eta\,\eta\,\rp\,E^{\scriptstyle{(1)}}{}_{\Sigma\,\tilde{P}}\\
{\color{Magenta}T^{\scriptstyle{(1)}}{}_{\tilde{P}\,D\,\Omega}}\equiv\,T^{\scriptstyle{(1)}}{}_{\tilde{\bf{a}}\,\bf{\alpha}}{}^{\bf{\beta}}&=&D_{(\bf{\alpha}}\,E^{\scriptstyle{(1)}}{}^{\bf{\beta})}{}_{\tilde{\bf{a}}}\,+\,\frac{1}{4}\,\gamma_{\bf{de}\,\,\bf{\alpha}}{}^{\bf{\beta}}\,E^{\scriptstyle{(1)}}{}^{\bf{de}}{}_{\tilde{\bf{a}}}\equiv\,D_{(D}\,E^{\scriptstyle{(1)}}{}_{\Omega)\tilde{P}}\,+	\,\varepsilon\,\gamma\,\rp\,E^{\scriptstyle{(1)}}{}_{\Sigma\tilde{P}}\\
{\color{Magenta}T^{\scriptstyle{(1)}}{}_{P\,D\,\tilde{\Omega}}}\equiv\,T^{\scriptstyle{(1)}}{}_{\bf{a}\,\bf{\alpha}}{}^{\tilde{\bf{\beta}}}&=&D_{[\bf{a}}\,E^{\scriptstyle{(1)}}{}_{\bf{\alpha}]}{}^{\tilde{\bf{\beta}}}\,+\,2\,\gamma_{\bf{a}\,\,\bf{\alpha}\,\bf{\epsilon}}\,E^{\scriptstyle{(1)}}{}^{\bf{\epsilon}\,\tilde{\bf{\beta}}}\equiv\,D_{[P}\,E^{\scriptstyle{(1)}}{}_{D]\tilde{\Omega}}\,+\,\gamma\,\rp\,E^{\scriptstyle{(1)}}{}_{\Omega\tilde{\Omega}}\\
{\color{Magenta}T^{\scriptstyle{(1)}}{}_{\tilde{D}\,\tilde{D}\,\Sigma}}\equiv T^{\scriptstyle{(1)}}{}_{\tilde{\bf{\alpha}}\,\tilde{\bf{\beta}}}{}^{\bf{ab}}&=&D_{(\tilde{\bf{\alpha}}}\,E^{\scriptstyle{(1)}}{}_{\tilde{\bf{\beta}})}{}^{\bf{ab}}\,+\,2\,\gamma^{\tilde{\bf{e}}}{}_{\tilde{\bf{\alpha}}\,\tilde{\bf{\beta}}}\,E^{\scriptstyle{(1)}}{}_{\tilde{\bf{e}}}{}^{\bf{ab}}\equiv\,D_{(\tilde{D}}\,E^{\scriptstyle{(1)}}{}_{\tilde{D})\Sigma}\,+\,\gamma\,\rp\,E^{\scriptstyle{(1)}}{}_{\tilde{P}\Sigma}
\end{rcases}
\ea
From the {(\ref{unmx})} is evident that by putting ${\color{Magenta}T^{\scriptstyle{(1)}}{}_{\tilde{P}\,P\,P}}\,=\,0$ we can determine $E^{\scriptstyle{(1)}}{}_{\Sigma\,\tilde{P}}$ in terms of $E^{\scriptstyle{(1)}}{}_{P\,\tilde{P}}$ and so $E^{\scriptstyle{(1)}}{}_{D\,\tilde{D}}$. Equivalently we can obtain that fixing either of ${\color{Magenta}T^{\scriptstyle{(1)}}{}_{\tilde{P}\,D\,\Omega}}$ or ${\color{Magenta}T^{\scriptstyle{(1)}}{}_{D\,D\,\tilde{\Sigma}}}\,$. By putting ${\color{Magenta}T^{\scriptstyle{(1)}}{}_{P\,D\,\tilde{\Omega}}}\,=\,0$ we can determine $E^{\scriptstyle{(1)}}{}_{\Omega\,\tilde{\Omega}}$ in terms of $E^{\scriptstyle{(1)}}{}_{D\,\tilde{\Omega}}$ and $E^{\scriptstyle{(1)}}{}_{P\,\tilde{\Omega}}$ and so again in $E^{\scriptstyle{(1)}}{}_{D\,\tilde{D}}$. The dimension $1$ mixed constraints give:
\ba
E^{\scriptstyle{(1)}}{}_{\Sigma\,\tilde{P}}\equiv\,E^{\scriptstyle{(1)}}{}^{\bf{bc}}{}_{\tilde{\bf{a}}}&=&\eta^{\bf{b\,d}}\,\eta^{\bf{c\,e}}\,D_{[\bf{d}}\,E^{\scriptstyle{(1)}}{}_{\bf{e}]\,\tilde{\bf{a}}}\,\equiv\,\eta\eta\,\rp\,D_{[P}\,E^{\scriptstyle{(1)}}{}_{P]\,\tilde{P}}\\
\label{kremik}
E^{\scriptstyle{(1)}}{}_{\Omega\,\tilde{\Omega}}\equiv\,E^{\scriptstyle{(1)}}{}^{\bf{\alpha}\,\tilde{\bf{\beta}}}&=&\frac{1}{6}\,\gamma^{\bf{a}\,\,\bf{\alpha}\,\bf{\epsilon}}\,D_{[\bf{a}}\,E^{\scriptstyle{(1)}}{}_{\bf{\epsilon}]}{}^{\tilde{\bf{\beta}}}\equiv\gamma\,\rp\,D_{[P}\,E^{\scriptstyle{(1)}}{}_{D]\tilde{\Omega}}
\ea

The dimension $1$ constraints could be viewed also form another perspective. For that we would need to borrow the expression for the \textit{Cartan-Killing} metric $K_{\As\,\Bs}$ that is discussed in section \ref{tvaroh}. The  expression for the linearised \textit{Cartan-Killing} metric:
\be
\label{smutnysom}
K^{\scriptsize{(1)}}{}_{\As\,\Bs}\,\equiv\,\frac{1}{2}\,f_{(\,\As\,|\,\Cs}{}^{\Ds}\,T^{{\scriptstyle (1)}}{}_{\Bs\,]\,\Ds}{}^{\Cs}
\ee
taking the (\ref{smutnysom}) for $\As,\,\Bs\,\in\,\{\,\alpha,\,\tilde{\beta}\,\}$ we will get:
\ba
K^{\scriptsize{(1)}}{}_{\alpha\,\beta}\,\propto\,\varepsilon_{\alpha\,\beta}\,B,\,\,\,\,K^{\scriptsize{(1)}}{}_{\tilde{\alpha}\,\tilde{\beta}}\,\propto\,\varepsilon_{\tilde{\alpha}\,\tilde{\beta}}\,\tilde{B},\,\,\,\,K^{\scriptsize{(1)}}{}_{{\alpha}\,\tilde{\beta}}
\ea
then using the exercise XA$2$.$6$ in \cite{fields} we could write the dimension $1$ constraints as:
\ba
T^{\scriptsize{(1)}}{}_{\bf{a}\,\bf{b}\,\bf{c}}\,\propto\,\varepsilon_{\bf{a}\,\bf{b}\,\bf{c}}\,\varepsilon^{\alpha\,\beta}\,K^{\scriptsize{(1)}}{}_{\beta\,\alpha}\,\,,&&T^{\scriptsize{(1)}}{}_{\alpha\,\beta}{}^{\bf{ab}}\,\propto\,\gamma^{\bf{ab}}{}_{\alpha\,\beta}\,\varepsilon^{\epsilon\,\sigma}\,K^{\scriptsize{(1)}}{}_{\sigma\,\epsilon}\\
T^{\scriptsize{(1)}}{}_{\bf{a}\,\alpha}{}^{\beta}\,\propto\,\gamma_{\bf{a}\,}{}_{\bf{\alpha}}^{\bf{\beta}}\,\varepsilon^{\epsilon\,\sigma}\,K^{\scriptsize{(1)}}{}_{\sigma\,\epsilon}\,\,,&&T^{\scriptstyle{(1)}}{}_{\bf{a}\,\tilde{\bf{\alpha}}}{}^{\bf{\beta}}\,\propto\,\gamma_{{\bf{a}}}{}^{{\beta}\,{\epsilon}}\,K^{\scriptsize{(1)}}{}_{\epsilon\,\tilde{\alpha}}\\
\label{veselysom}
T^{\scriptstyle{(1)}}{}_{\tilde{\bf{\alpha}}\,\beta}{}^{\bf{ab}}\,\propto\,\gamma^{\bf{ab}}{}_{\beta}{}^{\epsilon}\,K^{\scriptsize{(1)}}{}_{\epsilon\,\tilde{\alpha}}\,\,,&&T^{\scriptstyle{(1)}}{}_{\bf{a}\,\bf{\alpha}}{}^{\tilde{\bf{\beta}}}\,\propto\,\gamma_{{\bf{a}}}{}_{\,\,{\alpha}\,{\epsilon}}\,K^{\scriptsize{(1)}}{}^{\,\,\,\epsilon\,\tilde{\beta}}
\ea
Remaining dimension $1$ torsions have to be $0$ since we do not have appropriate nonzero \textit{Cartan-Killing} metric. We also put second torsion of (\ref{veselysom}) to $0$. Since that does not produce any differential constraints and fixes $\Eone{}_{\Omega\,\tilde{\Omega}}$, see (\ref{kremik}). Moreover in the spirit of the exercise XA$2$.$6$ in \cite{fields}, we can identify $(\,K^{\scriptsize{(1)}}{}_{{\alpha}\,\tilde{\beta}},\,B,\,\tilde{B}\,)$ with a $SO(\,3,\,3\,)$ vector $G^{\alpha\,\beta}\,=\,(\,G^{a},\,B,\,\bar{B}\,)$ in $SL(\,4\,)$ notation (form the $\sN=1$ supergravity).  
\subsection{\texorpdfstring{$\tilde{T}\,=\,0$}{\tilde{T}\,=\,0} constraints}
In the previous subsections we discovered that all the vielbeins (mixed and unmixed) (except of $\Eone{}_{\Omega\,\Sigma}$ and $\Eone{}_{\Sigma\,\Sigma}$) could be determined in terms of $\Eone{}_{P\Omega}$ and $\Eone{}_{D\tilde{D}}$. We need further constraint to relate those two undetermined vielbeins. We are following article \cite{warren}. There a new torsion was introduced. It came from the requirement of partial integration also in the presence of the new integration measure $\phi^2$ (dilaton). Following \cite{warren} the new torsion is:
\be
\label{astalos}
\tilde{T}_{\As}\,\defeq\,\phi^{2}\,\overleftarrow{\nabla}_{\As}\,\phi^{-\,2}
\ee
where $\nabla_{\As}\,=\,E_{\As}{}^{\Ms}\,D_{\Ms}$. The torsion (\ref{astalos}) should vanish, so we get the $\tilde{T}$ torsion constraint: $\tilde{T}_{\As}\,=\,0$. We are interested just in the first order part of $\tilde{T}_{\As}$:
\ba
\label{label}
\tilde{T}_{\As}\,=\,0\,+\,\tilde{T}^{\scriptstyle{(1)}}{}_{\As}\,+\,\Os(\,E^{\scriptstyle{(2)}}\,)&\Rightarrow&\tilde{T}^{\scriptstyle{(1)}}{}_{\As}\,=\,D^{\Bs}\,\Eone{}_{\Bs\,\As}\,+\,2\,D_{\As}\,\phi^{\scriptstyle{(1)}}\\
\text{where}&&\phi\,=\,1\,+\,\phi^{\scriptstyle{(1)}}\,+\,\mathcal{O}\,(\phi^{\scriptstyle{(2)}})\nonumber
\ea
The relation $\tilde{T}^{\scriptstyle{(1)}}{}_{S}\,=\,0$ gives $D_{S}\,\phi^{\scriptstyle{(1)}}\,=\,0$. Using $\tilde{T}^{\scriptsize{(1)}}{}_{D}\,=\,0$ we get the relation:
\ba
\label{ta}
\frac{1}{4}\,\varepsilon_{\bf{a\,b\,c}}\,\gamma^{\bf{c}\,}{}_{\bf{\alpha}}^{\bf{\beta}}\,\Eone{}^{\bf{ab}}{}_{\bf{\beta}}\,=\,2\,\gamma_{\bf{a}\,\bf{\alpha}\,\bf{\beta}}\,\Eone{}^{\,\bf{\beta}\,\bf{a}}\,&=&\,D^{\tilde{\bf{\beta}}}\,\Eone{}_{\tilde{\bf{\beta}}\,\alpha}\,+\,D^{\tilde{\bf{a}}}\,\Eone{}_{\tilde{\bf{a}}\,\alpha}\,-\,D_{\tilde{\bf{\beta}}}\,\Eone{}^{\,\tilde{\bf{\beta}}}{}_{\alpha}\,-\,2\,D_{\alpha}\,\phi^{\scriptstyle{}(1)}\nonumber\\
\gamma\,\rp\,\Eone{}_{\Omega P}&=&\,D_{\tilde{\Omega}}\,\Eone{}_{\tilde{D}D}\,+\,D_{\tilde{P}}\,\Eone{}_{\tilde{P}D}\,-\,D_{\tilde{D}}\,\Eone{}_{\tilde{\Omega}D}\,\\
&&-\,D_{D}\,\phi^{\scriptstyle{(1)}}\nonumber
 \ea
Where we used the results of table \texttt{(\ref{hrncek1})}. Using the table \texttt{(\ref{hrncek2})} for $\Eone{}_{\tilde{\bf{a}}\,\alpha}$ and $\Eone{}^{\,\tilde{\bf{\beta}}}{}_{\alpha}$ we have the relation between $\Eone{}_{P\Omega}$ and $\Eone{}_{D\tilde{D}}$ and linearised dilaton $\phi^{\scriptstyle{(1)}}$:
\ba
\label{juno}
\gamma\,\rp\,\Eone{}_{\Omega\,P}\,\equiv\,2\,\gamma^{\bf{a}}{}_{\bf{\alpha\,\beta}}\,\Eone{}^{\,\bf{\beta}}{}_{\bf{a}}&=&-\,\frac{1}{3}\,\gamma^{\tilde{\bf{a}}}{}^{\,\,\tilde{\bf{\beta}}\,\tilde{\bf{\epsilon}}}\,[\,D_{\tilde{\bf{a}}},\,D_{\tilde{\bf{\beta}}}\,]\,\Eone{}_{\tilde{\bf{\epsilon}}\,\bf{\alpha}}\,+\,\frac{1}{2}\,\gamma^{\tilde{\bf{a}}}{}^{\,\,\tilde{\bf{\beta}}\,\tilde{\bf{\epsilon}}}\,D_{\tilde{\bf{a}}}\,D_{\tilde{\bf{\beta}}}\,\Eone{}_{\tilde{\bf{\epsilon}}\,\alpha}\\
&&-\,2\,D_{\alpha}\,\phi^{\scriptstyle{(1)}}\nonumber\\
&\equiv&-\,\gamma\,\rp\,[\,D_{\tilde{P}},\,D_{\tilde{D}}]\,\Eone{}_{\tilde{D}\,D}\,+\,\gamma\,\rp\,D_{\tilde{P}}\,D_{\tilde{D}}\,\Eone{}_{\tilde{D}\,D}\,-\,D_{D}\,\phi^{\scriptstyle{(1)}}\nonumber
\ea
We notice that result (\ref{juno}) is exactly the right combination in order to express $B$ from (\ref{fraktura}) in terms of $\Eone{}_{\tilde{D}\,D}$ and $\phi^{\scriptstyle{(1)}}$. This will be used in next sections:
\ba
\label{kruotka}
B&=&\,-\,\frac{1}{\vartheta\,+\,6\,\zeta}\,\varepsilon^{\bf{\nu}\,\bf{\alpha}}\,D_{\bf{\nu}}\,{\Big[}\,\gamma^{\tilde{\bf{a}}\,\,\tilde{\bf{\beta}}\,\tilde{\bf{\epsilon}}}\,{\Big(}-\,\frac{1}{6}[\,D_{\tilde{\bf{a}}},\,D_{\tilde{\bf{\beta}}}]\,+\,\frac{1}{4}\,D_{\tilde{\bf{a}}}\,D_{\tilde{\bf{\beta}}}\,{\Big)}\,\Eone{}_{\tilde{\bf{\epsilon}}\,\bf{\alpha}}\,-\,D_{\alpha}\,\phi^{\scriptstyle{(1)}}\,{\Big]}\\
&\equiv&\varepsilon\,\rp\,D_{D}\,{\Big[}\,\gamma\,\rp\,{\Big(}\,[\,D_{\tilde{P}},\,D_{\tilde{D}}\,]\,-\,D_{\tilde{P}}\,D_{\tilde{D}}{\Big)}\,\Eone{}_{\tilde{D}\,D}\,-\,D_{D}\,\phi^{\scriptstyle{(1)}}\,{\Big]}\nonumber
\ea

Using the relation $\tilde{T}^{\scriptsize{(1)}}{}_{P}\,=\,0$ and similar steps we get:
\ba
\label{tata}
{ -\,2\,\varepsilon_{\bf{a}\,\bf{b\,c}}\,\gamma^{\bf{b}}{}^{\,\bf{\alpha}}_{\,\bf{\epsilon}}\,D_{\bf{\alpha}}\,\Eone{}^{\,\,\bf{\bf{\epsilon}}\,\bf{c}}\,+\,2\,D_{\bf{\alpha}}\,\Eone{}^{\bf{\alpha}}{}_{\bf{a}}\,-\,4\,D_{\bf{a}}\,\phi^{(1)}}&{ =}&{-\,\frac{1}{3}\gamma^{\tilde{\bf{a}}}{}^{\,\,\tilde{\bf{\beta}}\,\tilde{\bf{\epsilon}}}\,[\,D_{\tilde{\bf{a}}},\,D_{\tilde{\bf{\beta}}}\,]\,\gamma_{\bf{a}}{}^{\bf{\sigma}\,\bf{\alpha}}\,D_{\bf{\sigma}}\,\Eone{}_{\tilde{\bf{\epsilon}}\,\bf{\alpha}}}\\
&&+\,\frac{1}{2}\,\gamma^{\tilde{\bf{a}}}{}^{\,\,\tilde{\bf{\beta}}\,\tilde{\bf{\epsilon}}}\,D_{\tilde{\bf{a}}}\,D_{\tilde{\bf{\beta}}}\,\gamma_{\bf{a}}{}^{\bf{\sigma}\,\bf{\alpha}}\,D_{\bf{\sigma}}\,\Eone{}_{\tilde{\bf{\epsilon}}\,\alpha}\nonumber\\
{\Big{(}}\,\text{using (\ref{juno})}\,{\Big{)}}&{=}&{ \gamma_{\bf{a}}{}^{\bf{\sigma}\,\bf{\alpha}}\,D_{\bf{\sigma}}{\Big(}\,-\,2\,\gamma^{\bf{b}}{}_{\bf{\alpha\,\beta}}\,\Eone{}^{\,\bf{\beta}}{}_{\bf{b}}\,\,\,\,\,{\Big)}}\nonumber\\
{ -\,2\,\varepsilon_{\bf{a}\,\bf{b\,c}}\,\gamma^{\bf{b}}{}^{\,\bf{\alpha}}_{\,\bf{\epsilon}}\,D_{\bf{\alpha}}\,\Eone{}^{\,\,\bf{\bf{\epsilon}}\,\bf{c}}\,+\,2\,D_{\bf{\alpha}}\,\Eone{}^{\bf{\alpha}}{}_{\bf{a}}\,-\,4\,D_{\bf{a}}\,\phi^{(1)}}&{=}&{ -\,2\,\varepsilon_{\bf{a}\,\bf{b\,c}}\,\gamma^{\bf{b}}{}^{\,\bf{\alpha}}_{\,\bf{\epsilon}}\,D_{\bf{\alpha}}\,\Eone{}^{\,\,\bf{\bf{\epsilon}}\,\bf{c}}\,+\,2\,D_{\bf{\alpha}}\,\Eone{}^{\bf{\alpha}}{}_{\bf{a}}}\nonumber\\
&&-\,4\,D_{\bf{a}}\,\phi^{(1)}\nonumber
\ea
So, from relation $\tilde{T}^{\scriptstyle{(1)}}{}_{P}\,=\,0$ we will get no new constraints. 

From relations $\tilde{T}^{\scriptstyle{(1)}}{}_{\Omega}\,=\,0$ and $\tilde{T}^{\scriptstyle{(1)}}{}_{\Sigma}\,=\,0$ we will get some constraints on unfixed (and unused) vielbeins $\Eone{}_{\Sigma\,\Omega}$ and $\Eone{}_{\Sigma\,\Sigma}$. 

\section{Cartan-Killing metric and field equations}

\subsection{Cartan-Killing metric}
\label{tvaroh}
Having the Lie algebra ${\mathcal{G}}$, one can define a symmetric bilinear form:
\ba
\label{ck1}
K\,(\,X,\,Y\,)\,\defeq\,\frac{1}{x_{\lambda}}\,\mbox{Tr}\,(\,\mbox{ad}_{X}\,\mbox{ad}_{Y}\,)\,&\equiv&\,\frac{1}{x_{\lambda}}\,\Bra{E^{i}}\mbox{ad}_{X}\,\mbox{ad}_{Y}\,\Ket{E_{i}}\\
\mbox{where}&&X,\,Y\,\in\,{\mathcal G}\,\,\,\mbox{and}\,\,x_{\lambda}\,\equiv\,\mbox{Dynkin index}\nonumber\\
\mbox{and}&&E_{i},\,E^{j}\,\in\,{\mathcal{G}}\,\mbox{and} \,{\mathcal{G}^{*}}\nonumber
\ea
then for $X,\,Y\,\in$ basis of ${\mathcal G}$:
\ba
\label{ck2}
K\,(\,E_{i},\,E_{j}\,)\,\equiv\,&K_{ij}\,=\,&\,\frac{1}{x_{{\footnotesize \mbox{ad}}}}\,f_{i\,m}{}^{n}\,f_{j\,n}{}^{m}\\
\mbox{where}\,\,\,f_{a\,b}{}^{c}&\mbox{are}&\mbox{struc. cons. of}\,\,\,{\mathcal{G}}\nonumber
\ea
The \textit{Cartan-Killing} metric has many important group theoretical properties. We are interested in it because the field equations for the background fields could be viewed as if the level of the (engineering) dimension $1$ of the (generalised) \textit{Cartan-Killing} metric takes its free value. To see that, we need to generalise the \textit{Cartan-Killing} metric (\ref{ck2}) to the case of the (inhomogenous) graded algebra (\ref{algebra}). We use the direct generalisation of the expression (\ref{ck2}) for the algebra (\ref{algebra}) in the presence of the background fields (vielbeins). In that case the structure constants are given by (\ref{torsion}). We get (the Dynkin index $x_{{\footnotesize \mbox{ad}}}\,=\,2$):
\ba
K_{{\scriptsize\mathcal A\,B}}\,=\,\frac{1}{2}\,T_{\As\,\Cs}{}^{\Ds}\,T_{\Bs\,\Ds}{}^{\Cs}
\ea
We are interested in linearised version of previous equation. Again we expand the vielbeins to the first order and get: 
\ba
\label{nufka}
K_{\As\,\Bs}\,&=&\,\frac{1}{2}\,f_{\As\,\Cs}{}^{\Ds}\,f_{\Bs\,\Ds}{}^{\Cs}\,+\,\underbrace{\frac{1}{2}\,f_{(\,\As\,|\,\Cs}{}^{\Ds}\,T^{{\scriptstyle (1)}}{}_{\Bs\,]\,\Ds}{}^{\Cs}}_{K^{\scriptstyle (1)}{}_{\As\,\Bs}}\,+\,{\mathcal O}\,(\,E^{\scriptstyle (2)}\,)\\
\mbox{where}\,\,\,\,\,\,\,T^{{\scriptstyle (1)}}{}_{\As\,\Bs\,\Cs}&\defeq&\frac{1}{2}\,D_{[\,\As}\,E^{{\scriptstyle (1)}}{}_{\Bs\,\Cs\,)}\,+\,\frac{1}{2}\,\Eone{}_{[\As}{}^{\Ms}\,f_{\Ms|\,\Bs\,\Cs\,)}\nonumber
\ea
\subsection{Field equations}
After imposing all the constraints we have found that everything could be expressed in terms of $\Eone{}_{P\,\Omega}$ and $\Eone{}_{D\,\tilde{D}}$. The gamma ``trace" part of $\Eone{}_{P\,\Omega}$ is related directly to $\Eone{}_{D\,\tilde{D}}$ by (\ref{juno}). Therefore we want equation of motion for the field $\Eone{}_{D\,\tilde{D}}$. 

We start with some action $S$ and vary it with respect to vielbein $E^{D\,D}$ and put it to the zero, i.e. $\rfrac{\delta\,}{\delta\,E^{\scriptsize{DD}}}\,S\,=\,0$. The variation produces the dimension $1$ antisymmetric tensor. On the other hand in the previous subsection we have seen that $K_{D\,D}$ is the canonical antisymmetric dimension $1$ tensor. Therefore we can impose the equations of motion:
\be
\label{eom}
\frac{\delta}{\delta\,E^{\scriptsize{DD}}}\,S\,\equiv\,K_{D\,D}\,=\,0
\ee
For the vielbein $E_{\tilde{\alpha}\,\beta}$ we produce the following equations:
\be
K_{\tilde{\alpha}\,\beta}\,=\,0\,\Rightarrow\,K^{\scriptsize(1)}{}_{\tilde{\alpha}\,\beta}\,=\,0
\ee
Plugging the definitions of structure constants and linearised torsions (note that only the combination of dimension $1$ torsions is present, since lower dimensional $T^{\scriptsize(1)}$ torsions are all set to zero):
\be
\label{Prague}
\,-\,2\,\gamma^{\tilde{\bf{m}}}{}_{\tilde{\bf{\alpha}}\,\tilde{\bf{\nu}}}\,T^{\scriptsize(1)}{}_{\bf{\beta}\,\tilde{\bf{m}}}{}^{\tilde{\bf{\nu}}}\,+\,2\,\gamma^{{\bf{m}}}{}_{{\bf{\beta}}\,{\bf{\nu}}}\,T^{\scriptsize(1)}{}_{\tilde{\bf{\alpha}}\,{\bf{m}}}{}^{{\bf{\nu}}}\,-\,\frac{1}{8}\,\varepsilon_{\tilde{\bf{a}}\,\tilde{\bf{b}}\,\tilde{\bf{c}}}\,\gamma^{\tilde{\bf{c}}}{}^{\,\,\tilde{\nu}}_{\,\,\tilde{\alpha}}\,T^{\scriptsize(1)}{}_{\beta\,\tilde{\nu}}{}^{\tilde{\bf{a}}\tilde{\bf{b}}}\,+\,\frac{1}{8}\,\varepsilon_{{\bf{a}}\,{\bf{b}}\,{\bf{c}}}\,\gamma^{{\bf{c}}}{}^{\,\,{\nu}}_{\,\,{\beta}}\,T^{\scriptsize(1)}{}_{\tilde{\alpha}\,{\nu}}{}^{{\bf{a}}{\bf{b}}}\,=\,0
\ee 
To simplify (\ref{Prague}) we can use one of the linearised (super)Bianchi identities that relates $T^{\scriptsize(1)}{}_{\tilde{\alpha}\,\beta}{}^{\bf{a}\bf{b}}\,\equiv\,T^{\scriptsize(1)}{}_{\tilde{D}\,D\,\Sigma}$ with $T^{\scriptsize(1)}{}_{\tilde{\alpha}}{}^{\,\bf{a}\,\nu}\,\equiv\,T^{\scriptsize(1)}{}_{\tilde{D}\,P\,\Omega\,}$:
\be
\label{Rome}
T^{\scriptsize(1)}{}_{\tilde{\alpha}\,\beta}{}^{\bf{ab}}\,=\,2\,\gamma^{[\bf{a}}{}_{\beta\,\nu}\,T^{\scriptsize(1)}{}^{\,\,\bf{b}]}{}_{\tilde{\alpha}}{}^{\nu}
\ee
Doing that we can see that the field equation (\ref{Prague}) becomes:
\be
\label{Zilina}
-\gamma^{\tilde{\bf{m}}}{}_{\tilde{\bf{\alpha}}\,\tilde{\bf{\nu}}}\,T^{\scriptsize(1)}{}_{\bf{\beta}\,\tilde{\bf{m}}}{}^{\tilde{\bf{\nu}}}\,+\,\gamma^{{\bf{m}}}{}_{{\bf{\beta}}\,{\bf{\nu}}}\,T^{\scriptsize(1)}{}_{\tilde{\bf{\alpha}}\,{\bf{m}}}{}^{{\bf{\nu}}}\,=\,0
\ee
Using the explicit knowledge of $T^{\scriptsize (1)}{}_{\tilde{\alpha}\,\bf{m}}{}^{\nu}$ from the table {\texttt{(\ref{hrncek2})}} and also the result of the $\tilde{T}_{D}$ constraint (\ref{juno}). The (\ref{Zilina}) could be rewritten as the differential equation just for the vielbein $\Eone{}_{\tilde{D}\,D}\,\equiv\,\Eone{}_{\tilde{\alpha}\,\beta}$. For the completeness we give the e.o.m. for the vielbein $\Eone{}_{\tilde{\alpha}\,\beta}$:
\ba
&\Big{[}\delta_{\tilde{\alpha}}^{\tilde{\sigma}}\Big{(}\,-\,\frac{1}{2}\,\gamma^{\bf{m}\,\nu}_{\,\,\,\,\,\,\,\beta}\,D_{\bf{m}}\,D^2\,-\,\delta_{\beta}^{\nu}\,\Box\,-\,\gamma^{\,\,\,\,\,\nu}_{\bf{s}\,\,\beta}\,.\,(\,D_{\bf{m}}\,\times\,D_{\bf{a}}\,)\,-\,2\,\delta_{\beta}^{\nu}\,D^{\mu}\,D_{\mu}\,-\,2\,D^{\nu}\,D_{\beta}\,\Big{)}&\\
&+\frac{1}{2}\,\delta_{\beta}^{\nu}\,\epsilon^{\tilde{\sigma}\tilde{\epsilon}}\,D_{\tilde{\alpha}}\,\tilde{D}^2\,D_{\tilde{\epsilon}}\,-\,(\,\tilde{\alpha}\,\rightarrow\,\beta\,\text{and}\,\beta\,\rightarrow\,\tilde{\alpha}\,)\,\Big{]}\,\Eone{}_{\tilde{\sigma}\,\nu}=\,0&\nonumber\\
&\Big{[}\Big{(}\,\gamma\,\rp\,D_{P}\,D_{D}\,\rp\,D_{D}\,-\,D_{P}\,\rp\,D_{P}\,-\,\gamma\,\rp\,(\,D_{P}\,\rp\,D_{P}\,)\,-\,D_{\Omega}\,\rp\,D_{D}\,-\,D_{\Omega}\,D_{D}\,\Big{)}&\\
&+\,D_{\tilde{D}}\,D_{\tilde{D}}\,\rp\,D_{\tilde{D}}\,D_{\tilde{D}}\,-\,(\,\tilde{D}\,\leftrightarrow\,D\,)\,\Big{]}\,\Eone{}_{\tilde{D}\,D}=0&\nonumber
\ea
where $\Box\,\equiv\,\eta^{a\,b}\,D_{a}\,D_{b}$ and $\gamma^{\,\,\,\,\,\nu}_{\bf{s}\,\,\beta}\,.\,(\,D_{\bf{m}}\,\times\,D_{\bf{a}}\,)\,\equiv\,\varepsilon^{\bf{s}\,\bf{m}\,\bf{a}}\,\gamma^{\,\,\,\,\,\nu}_{\bf{s}\,\,\beta}\,D_{\bf{m}}\,D_{\bf{a}}\,\propto\,D_{\Sigma}$.

The remaining equations are obtained by variation of the $S$ with respect to $E^{\alpha\,\beta}$ and $E^{\tilde{\alpha}\,\tilde{\beta}}$. We get:
\ba
\label{Brussels}
K_{\alpha\,\beta}\,=\,K^{\scriptsize{(1)}}{}_{\alpha\,\beta}\,=\,0\,\,\,&\text{and}&\,\,\,K_{\tilde{\alpha}\,\tilde{\beta}}\,=\,K^{\scriptsize{(1)}}{}_{\tilde{\alpha}\,\tilde{\beta}}\,=\,0\\
&\text{where}&\nonumber\\
\label{Bratislava}
K^{\scriptsize{(1)}}{}_{\alpha\,\beta}\,\propto\,\varepsilon_{\alpha\,\beta}\,B\,&&K^{\scriptsize{(1)}}{}_{\tilde{\alpha}\,\tilde{\beta}}\,\propto\,\varepsilon_{\tilde{\alpha}\,\tilde{\beta}}\,\tilde{B}
\ea
Equations (\ref{Brussels}) and (\ref{Bratislava}) could be rewritten in a different way:
\be
\label{Lenuska}
B\,+\,\tilde{B}\,=\,0\,\,\,\text{and}\,\,\,B\,-\,\tilde{B}\,=\,0
\ee
where $B$ is given by eq. (\ref{kruotka}). Because the explicit structure of $B$ and $\tilde{B}$ is important for the next considerations we repeat it here:
\ba
\label{Gdansk}
{ B}&{ \propto}&{ \varepsilon^{\nu\,\alpha}\,D_{\nu}\,(\,D^{\tilde{\epsilon}}\,+\,\frac{1}{4}\,\gamma^{\tilde{\bf{a}}\,\,\tilde{\beta}\,\tilde{\epsilon}}\,D_{\tilde{\bf{a}}}\,D_{\tilde{\beta}}\,)\,E^{(1)}{}_{\tilde{\epsilon}\,\alpha}\,+\,\varepsilon^{\alpha\,\beta}\,D_{\beta}\,D_{\alpha}\,\phi^{(1)}}\\
&\equiv&\varepsilon\,\rp\,D_{D}\,(\,D_{\tilde{\Omega}}\,+\,\gamma\,\rp\,D_{\tilde{P}}\,D_{\tilde{D}}\,)\,E^{(1)}{}_{\tilde{D}\,D}\,+\,D_{D}\,\rp\,D_{D}\,\phi^{(1)}\nonumber\\
\label{ksnadG}
{ \tilde{B}}&{\scriptstyle \propto}&{\varepsilon^{\tilde{\nu}\,\tilde{\epsilon}}\,D_{\tilde{\nu}}\,(\,-\,D^{\alpha}\,+\,\frac{1}{4}\,\gamma^{\bf{a}\,\,\beta\,\alpha}\,D_{\bf{a}}\,D_{\beta}\,)\,E^{(1)}{}_{\tilde{\epsilon}\,\alpha}\,+\,\varepsilon^{\tilde{\alpha}\,\tilde{\beta}}\,D_{\tilde{\beta}}\,D_{\tilde{\alpha}}\,\phi^{(1)}\,}\\
&\equiv&\varepsilon\,\rp\,D_{\tilde{D}}\,(\,-\,D_{\Omega}\,+\,\gamma\,\rp\,D_{P}\,D_{D}\,)\,E^{(1)}{}_{\tilde{D}\,D}\,+\,{D}_{\tilde{D}}\,\rp\,{D}_{\tilde{D}}\,\phi^{(1)}\nonumber
\ea

To analyse the second terms in (\ref{Gdansk}) and (\ref{ksnadG}) we need the following identities:
\ba
\label{pipjaky}
\gamma^{\bf{a}\,\,\beta\,\alpha}\,D_{\bf{a}}\,D_{\beta}&=&4\,D^{\alpha}\,-\,\frac{1}{2}\,D^{2}\,\varepsilon^{\alpha\,\epsilon}\,D_{\epsilon}\,\equiv\,D_{\Omega}\,-\,(D_{D}\,\rp\,D_{D})\,\varepsilon\,D_{D}\\
\label{lenka}
\gamma^{\tilde{\bf{a}}\,\,\tilde{\beta}\,\tilde{\alpha}}\,D_{\tilde{\bf{a}}}\,D_{\tilde{\beta}}&=&-\,4\,D^{\tilde{\alpha}}\,+\,\frac{1}{2}\,\tilde{D}^{2}\,\varepsilon^{\tilde{\alpha}\,\tilde{\epsilon}}\,D_{\tilde{\epsilon}}\,\equiv\,-\,D_{\tilde{\Omega}}\,+\,(D_{\tilde{D}}\,\rp\,D_{\tilde{D}})\,\varepsilon\,D_{\tilde{D}}
\ea 
where $D^2\,=\,\varepsilon^{\beta\,\alpha}\,D_{\alpha}\,D_{\beta}\,\equiv\,D_{D}\,\rp\,D_{D}$ (similarly for $\tilde{D}^{2}$). 

Using (\ref{pipjaky}) and (\ref{lenka}) we get:
\ba
{B}&{\propto}&{-\,\frac{1}{8}\,\tilde{D}^{2}\,(\,\varepsilon^{\alpha\,\nu}\,\varepsilon^{\tilde{\epsilon}\,\tilde{\sigma}}\,D_{\nu}\,D_{\tilde{\sigma}}\,E^{(1)}{}_{\tilde{\epsilon}\,\alpha}\,)\,+\,D^{2}\,\phi^{(1)}\,}\\
&\equiv&-\,{D}_{\tilde{D}}\,\rp\,{D}_{\tilde{D}}\,(\,D_{D}\,D_{\tilde{D}}\,)\,\rp\,\Eone{}_{\tilde{D}\,D}\,+\,D_{D}\,\rp\,D_{D}\,\phi^{(1)}\nonumber\\
{\tilde{B}}&{\propto}&{-\,\frac{1}{8}\,D^{2}\,(\,\varepsilon^{\alpha\,\nu}\,\varepsilon^{\tilde{\epsilon}\,\tilde{\sigma}}\,D_{\nu}\,D_{\tilde{\sigma}}\,E^{(1)}{}_{\tilde{\epsilon}\,\alpha}\,)\,+\,\tilde{D}^{2}\,\phi^{(1)}}\\
&\equiv&-\,{D}_{D}\,\rp\,{D}_{D}\,(\,D_{D}\,D_{\tilde{D}}\,)\,\rp\,\Eone{}_{\tilde{D}\,D}\,+\,{D}_{\tilde{D}}\,\rp\,{D}_{\tilde{D}}\,\phi^{(1)}\nonumber
\ea
Then the first of (\ref{Lenuska}) becomes the equation:
\be
\label{love}
{ 0\,=\,(\,D^{2}\,+\,\tilde{D}^{2}\,)\,(\,-\,\frac{1}{8}\,\varepsilon^{\alpha\,\nu}\,\varepsilon^{\tilde{\epsilon}\,\tilde{\sigma}}\,D_{\nu}\,D_{\tilde{\sigma}}\,E^{(1)}{}_{\tilde{\epsilon}\,\alpha}\,+\,\phi^{(1)}\,)}
\ee
We can rewrite (\ref{love}) using a new field $V$:
\be
\label{Ftrain}
{ (\,D^{2}\,-\,\tilde{D}^{2}\,)\,V\,\eqdef\,(\,-\,\frac{1}{8}\,\varepsilon^{\alpha\,\nu}\,\varepsilon^{\tilde{\epsilon}\,\tilde{\sigma}}\,D_{\nu}\,D_{\tilde{\sigma}}\,E^{(1)}{}_{\tilde{\epsilon}\,\alpha}\,+\,\phi^{(1)}\,)}
\ee
Using this definition the (\ref{love}) could be written as: 
\be
\label{peace}
0\,=\,(\,D^{2}\,+\,\tilde{D}^{2}\,)\,(\,D^{2}\,-\,\tilde{D}^{2}\,)\,V
\ee
The operator $(\,D^{2}\,+\,\tilde{D}^{2}\,)\,(\,D^{2}\,-\,\tilde{D}^{2}\,)$ is acting on the scalar field $V$. It can be rewritten in a nicer form:
\ba
\label{Etrain}
(\,D^{2}\,+\,\tilde{D}^{2}\,)\,(\,D^{2}\,-\,\tilde{D}^{2}\,)\,V&=&4\,(\Box\,-\,D_{\nu}\,D^{\nu}\,+\,D^{\nu}\,D_{\nu}\,-\,(\,\tilde{\Box}\,-\,D_{\tilde{\nu}}\,{D}^{\tilde{\nu}}\,+\,D^{\tilde{\nu}}\,D_{\tilde{\nu}}\,)\,)\,V\nonumber\\
&\equiv&4\,D^{\As}\,D_{\As}\,V
\ea
Therefore the first equation of (\ref{Lenuska}) could be rewritten as:
\be
\label{Gtrain}
D^{\As}\,D_{\As}\,V\,=\,0
\ee
and so (\ref{Gtrain}) is identically satisfied since it is just the strong constraint.

The second equation of (\ref{Lenuska}) becomes the e.o.m. for the $V$ field: 
\be
\label{Mtrain}
(\,D^{2}\,-\,\tilde{D}^{2}\,)^{2}\,V\,=\,0
\ee
\subsection{Field equations: Summary}
The field equations are summarised in the following table ${\color{black}\texttt{(\ref{hrncek4})}}$:
\begin{table}[H]
{\centering
\begin{tabular}{| ccc |}
\noalign{\vspace{.1in}}
\hline
$K^{\scriptsize(1)}{}_{\tilde{\alpha}\,\beta}\,=\,0$&$\Rightarrow$&$\Big{[}\delta_{\tilde{\alpha}}^{\tilde{\sigma}}\Big{(}\,-\,\frac{1}{2}\,\gamma^{\bf{m}\,\nu}_{\,\,\,\,\,\,\,\beta}\,D_{\bf{m}}\,D^2\,-\,\delta_{\beta}^{\nu}\,\Box\,-\,\gamma^{\,\,\,\,\,\nu}_{\bf{s}\,\,\beta}\,.\,(\,D_{\bf{m}}\,\times\,D_{\bf{a}}\,)$\\
&&$-\,2\,\delta_{\beta}^{\nu}\,D^{\mu}\,D_{\mu}\,-\,2\,D^{\nu}\,D_{\beta}\,\Big{)}\,+\,\frac{1}{2}\,\delta_{\beta}^{\nu}\,\epsilon^{\tilde{\sigma}\tilde{\epsilon}}\,D_{\tilde{\alpha}}\,\tilde{D}^2\,D_{\tilde{\epsilon}}\,$\\
&&$-\,(\,\tilde{\alpha}\,\rightarrow\,\beta\,\text{and}\,\beta\,\rightarrow\,\tilde{\alpha}\,)\,\Big{]}\,\Eone{}_{\tilde{\sigma}\,\nu}\,=\,0$\\
\hline
$K^{\scriptsize(1)}{}_{\alpha\,\beta}\,+\,K^{\scriptsize(1)}{}_{\tilde{\alpha}\,\tilde{\beta}}\,=\,0$&$\Rightarrow$&$D^{\As}\,D_{\As}\,V\,=\,0$\\
&where&$(\,D^{2}\,-\,\tilde{D}^{2}\,)\,V\,\eqdef\,(\,-\,\frac{1}{8}\,\varepsilon^{\alpha\,\nu}\,\varepsilon^{\tilde{\epsilon}\,\tilde{\sigma}}\,D_{\nu}\,D_{\tilde{\sigma}}\,E^{(1)}{}_{\tilde{\epsilon}\,\alpha}\,+\,\phi^{(1)}\,)$\\
\hline
$K^{\scriptsize(1)}{}_{\alpha\,\beta}\,-\,K^{\scriptsize(1)}{}_{\tilde{\alpha}\,\tilde{\beta}}\,=\,0$&$\Rightarrow$&$(\,D^{2}\,-\,\tilde{D}^{2}\,)^{2}\,V\,=\,0$\\
\hline
\end{tabular}
}
\caption{Field equations \protect\label{hrncek4}}
\end{table}

\section{Dilaton}
The result of the previous section gives the structure of the linear dilaton $\phi^{\scriptsize{(1)}}$, see table {\texttt{(\ref{hrncek4})}}. Using the relation (\ref{Ftrain}) we find the structure of the linear dilaton:
\be
\phi^{{(1)}}\,=\,\frac{1}{8}\,\varepsilon^{\alpha\,\nu}\,\varepsilon^{\tilde{\epsilon}\,\tilde{\sigma}}\,D_{\nu}\,D_{\tilde{\sigma}}\,E^{(1)}{}_{\tilde{\epsilon}\,\alpha}\,+\,(\,D^{2}\,-\,\tilde{D}^{2}\,)\,V\,\equiv\,\varepsilon\,\varepsilon\,D_{D}\,D_{\tilde{D}}\,\rp\,E^{(1)}{}_{D\,\tilde{D}}\,+\,(D_{D}\,\rp\,D_{D}\,-\,D_{\tilde{D}}\,\rp\,D_{\tilde{D}}\,)\,V
\ee
We notice that the structure of the linear dilaton matches the structure of the dilaton field obtained by compactifying the $4$D $\sN=1$ supergravity to $3$ dimensions, see section $7$.$2$.b in \cite{1001}. 
For the dilaton we can though impose the space-time action (after compactification of half of the dimensions, as usual in double field theory):
\be
S_{dil}\,\defeq\,\int\,d^{3}x\,d^{2}\theta\,\phi^{2}
\ee
where $\phi\,\approx\,1\,+\,\phi^{(1)}$. Moreover the cosmological constant could be added, then the action becomes:
\be
S_{dil}\,\defeq\,\int\,d^{3}x\,d^{2}\theta\,(\,\phi^{2}\,-\,\lambda\,V)
\ee

\section{Conclusion}
We outline results we have obtained: we started with the T-dual $\sN=2$ string theory, i.e. effective $\sN=2$ supergravity in $3$ dimensions. We knew that this theory should be equivalent to the theory obtained from the classical $\sN=1$ supergravity in $4$ dimensions. In this paper we first obtained the dimension $-\,1$ pre potential as the vielbein component $\Eone{}_{D\,\tilde{D}}\,\equiv\,\Eone{}_{\alpha\,\tilde{\beta}}$ and the dimension $-\,\frac{3}{2}$ unconstrained gauge parameter $\Lambda_{D}\,\equiv\,\Lambda_{\alpha}$  (also $\Lambda_{\tilde{D}}$) without solving any differential constraints. In the usual $4$ dimensional $\sN=1$ supergravity they appear only through their derivatives in objects of higher dimension after solving differential constraints, see section X.A.$1$ in \cite{fields}. We have also derived the structure of the $\sN=2$ supergravity in $3$ dimensions using the techniques of the T-dually extended superspace. In particular the structure of the linear dilaton $\phi$ has been derived. It matches the structure obtained from $4$D $\sN=1$ and its compactification, see section $7$.$9$ in \cite{1001} and \cite{graph}. This suggest that T-dualy extended superspace approach could be extended also to higher dimensional cases, see \cite{warren1}.  
\section*{Acknowledgment}
This work was supported in part by National Science Foundation Grant No. PHY-1316617.

\end{document}